\documentclass[twocolumn]{aastex631}

\usepackage{amsmath}
\usepackage{enumitem}
\usepackage{rotating}

\shorttitle{$\tau$-Herculid}
\shortauthors{Egal et al.}
\graphicspath{{./}{figures/}}

\newcommand{\degree}{{^{\circ}}}

\begin{document}

\title{Modelling the 2022 $\tau$-Herculid outburst}

\author[0000-0002-9572-1200]{Auriane Egal}
\affiliation{Planétarium Rio Tinto Alcan, Espace pour la Vie, 
4801 av. Pierre-de Coubertin, Montréal, Québec, Canada}
\affiliation{Department of Physics and Astronomy, 
University of Western Ontario, 
London, Ontario, N6A 3K7, Canada}
\affiliation{Institute for Earth and Space Exploration, 
University of Western Ontario, 
London, Ontario, N6A 3K7, Canada}
\affiliation{IMCCE, Observatoire de Paris, PSL Research University, CNRS, Sorbonne Universités,\\ UPMC Univ. Paris 06, Univ. Lille, France}

\author[0000-0002-1914-5352]{Paul A. Wiegert}
\affiliation{Department of Physics and Astronomy, 
University of Western Ontario, 
London, Ontario, N6A 3K7, Canada}
\affiliation{Institute for Earth and Space Exploration, 
University of Western Ontario, 
London, Ontario, N6A 3K7, Canada}

\author[0000-0001-6130-7039]{Peter G. Brown}
\affiliation{Department of Physics and Astronomy, 
University of Western Ontario, 
London, Ontario, N6A 3K7, Canada}
\affiliation{Institute for Earth and Space Exploration, 
University of Western Ontario, 
London, Ontario, N6A 3K7, Canada}

\author[0000-0003-4166-8704]{Denis Vida}
\affiliation{Department of Physics and Astronomy, 
University of Western Ontario, 
London, Ontario, N6A 3K7, Canada}
\affiliation{Institute for Earth and Space Exploration, 
University of Western Ontario, 
London, Ontario, N6A 3K7, Canada}



\begin{abstract}

 The $\tau$-Herculids (IAU shower number $\#61$ TAH) is a minor meteor shower associated with comet 73P/Schwassmann-Wachmann 3, a Jupiter-Family comet that disintegrated into several fragments in 1995. As a consequence of the nucleus break-up, possible increased meteor rates were predicted for 2022. On May 30-31, observation networks around the world reported two distinct peaks of TAH activity, around solar longitudes 69.02$\degree$ and 69.42$\degree$. This work examines the encounter conditions of the Earth with meteoroids ejected from 73P  during the splitting event and on previous perihelion passages. Numerical simulations suggest that the main peak observed in 2022 was caused by meteoroids ejected from the splitting nucleus with four times the typical cometary gas expansion speed. High-resolution measurements performed with the Canadian Automated Meteor Observatory indicate that these meteoroids are fragile, with estimated bulk densities of 250 kg/m$^3$.
 In contrast with the main peak, the first TAH activity peak in 2022 is best modelled with trails ejected prior to 1960. We find that ordinary cometary activity could have produced other TAH apparitions observed in the past, including in 1930 and 2017. The extension of our model to future years predicts significant returns of the shower in 2033 and 2049.

\end{abstract}



\section{Introduction} \label{sec:intro}

Comet 73P/Schwassmann-Wachmann 3 (hereafter 73P) is a Jupiter-family comet with a 5.4-year period observed for the first time in 1930.
Because of its faintness and the strong perturbations to its orbit caused by Jupiter, the comet was lost for a few decades after its discovery. Fortunately, the comet was recovered in 1979 and has been observed during most of its apparitions since 1990. Despite expectations of an unremarkable return of the comet in 1995, 73P experienced a major outburst in September of that year that increased its predicted brightness by a factor of 400 \citep{Rao2021}. Telescopic observations conducted in December 1995 revealed that the comet had split into at least four fragments, labeled 73P-A, B, C and D \citep{Bohnhardt1995}. The comet has undergone a series of subsequent disintegrations since then, resulting in the separation of several hundred fragments from the original nucleus \citep{Ishiguro2009}. 

Since its discovery, 73P has been linked to the $\tau$-Herculids meteor shower \citep{Nakamura1930}. The shower is designated as \#61 TAH by the Meteor Data Center\footnote{\url{https://www.ta3.sk/IAUC22DB/MDC2022/}}. TAH  displays are generally unimpressive, with little to no meteor activity recorded at each shower's return. With the exception of an outburst reported by a single observer in 1930 \citep{Nakamura1930}, the TAH are considered to be essentially inactive. However, the break-up of 73P's nucleus in 1995 raised expectations for enhanced TAH activity in the spring of 2022, produced by meteoroids released during the splitting process. 

Dynamical models of the meteoroids ejected by 73P in 1995 indicate that material released at typical cometary gas-drag ejection speeds \citep{Jones1995a} would not produce any strong TAH activity in 2022 \citep{Wiegert2005,Rao2021,Ye2022}. However, models assuming higher ejection speeds did predict enhanced TAH rates caused by the 1995 ejecta \citep[e.g.,][]{Luthen2001,Horii2008,Rao2021}. The hopes for a possible TAH outburst or storm in 2022 led meteor detection networks all around the world to organize observation campaigns to record the shower's return. 

As a result, several independent observers reported enhanced $\tau$-Herculid activity on May 31, 2022, reaching a zenithal hourly rate (ZHR) of 20 to 50 meteors per hour around 4h 15 UT \citep{Jenniskens2022,Ogawa2022,Vida2022,Weiland2022,Ye2022}. The shower was observed, among others, by the video cameras of the Global Meteor Network \citep[GMN,][]{Vida2021b}, the Canadian Automated Meteor Observatory \citep[CAMO,][]{Weryk2013,Vida2021}, the Cameras for Allsky Meteor Surveillance \citep[CAMS,][]{Jenniskens2011} and the International Meteor Organization Video Meteor Network (IMO VMN\footnote{http://www.imonet.org/imc13/meteoroids2013\_poster.pdf}). Many TAH meteors were detected by the instruments of a joint Australian-European airborne observation campaign\footnote{\url{https://www.imcce.fr/recherche/campagnes-observations/meteors/2022the}}. The shower was also recorded by the Canadian Meteor Orbit Radar \citep[CMOR,][]{Brown2008,Brown2010} and the International Project for Radio Meteor Observations \citep[IPRMO,][]{Ogawa2004}. Increased meteor activity was in addition reported by visual observers, as indicated in the IMO Visual Meteor DataBase (IMO VMDB\footnote{\url{https://www.imo.net/members/imo_live_shower?shower=TAH\&year=2022}, accessed in September 2022}). 

Recently, \cite{Ye2022} examined the encounter conditions in 2022 of meteoroids produced during the 1995 breakup of 73P. The authors explored different ejection scenarios for the mm and sub-millimeter class meteoroids released by the comet, increasing the  ejection speeds of the particles from 1 to 5 times the values predicted by the models of \cite{Whipple1951} and \cite{Crifo1997}. In all their scenarios, sub-millimeter particles (primarily radar meteors at TAH speeds) were found to intersect Earth's orbit in 2022, while mm-class meteoroids (optical meteors at TAH speeds) reached Earth only for ejection speeds from the nucleus exceeding 2.5 to 2.75 times \cite{Whipple1951}'s nominal gas-drag values. The best match with the observed ZHR and full width at half-maximum (FWMH) was found for ejection speeds reaching 4 to 5 times \cite{Whipple1951}'s. Such velocities are twice those determined for particles comprising 73P's trail by the Spitzer Space Telescope \citep{Vaubaillon2010}.

Centimeter-sized meteoroids, necessary to explain the existence of several bright TAH meteors observed in 2022\footnote{\color{blue!75!yellow}{https://www.meteornews.net/2022/08/24/a-meteor-outburst-caused-by-dust-from-comet-73p-schwassmann-wachmann-the-tau-herculids-a-visual-analysis/}}, were found unlikely to approach Earth in simulations by \cite{Ye2022}. The authors thus suggested that the brightest meteors observed could result from the disintegration of porous $\ge$cm dust aggregates, that followed trajectories similar to mm-sized particles.  

In this work, we present results of our modelling of the $\tau$-Herculids from 1930 onwards, calibrated on observations of the 2022 shower performed by CAMO, CMOR, and the GMN. As a first step, we determine the physical properties of two TAH meteoroids using CAMO's high-resolution optical measurements. We then examine the relative contribution of the meteoroids ejected prior to and during the 1995 break-up of 73P's nucleus in order to reproduce most of the shower features observed in 2022, such as radiant location and activity profile. Finally, we extend our model to future apparitions of the shower and forecast its activity until 2050. 

\section{2022 Tau Herculids} \label{sec:observations_2022}

\subsection{Physical properties}
The Canadian Automated Meteor Observatory (CAMO) consists of several optical instruments located at two sites in Southern Ontario, Canada. The sites, separated by 50 km, have optical instruments pointed to overlapping atmospheric regions to permit meteor triangulation. Among the instruments at each site is a high-resolution mirror tracking system. This system observes meteors through a telescope using Gen 3 image intensifiers coupled to a video camera, achieving high temporal cadence (100 frames per second) and meter-scale spatial resolution. Upon meteor detection in a separate wide-field camera, two mirrors are cued to track the meteor in real time. The mirrors then direct the meteor light into the narrow-field ($1.5^{\circ} \times 1.5^{\circ}$) stationary telescope \citep{Weryk2013, Vida2021}. 

On May 31, 2022 near the time of the peak of $\tau$-Herculids the CAMO mirror tracking system detected two shower members which were well observed at both stations. As shown in Appendix \ref{sec:physical_properties} Figure \ref{fig:tau_herc}, the high-resolution video shows clear evidence of extensive fragmentation and wake, consistent with a fragile meteoroid. 

Following the data reduction procedures described in \cite{Vida2021}, the position and the photometry of the meteor on each frame from each site were manually measured. The wide field photometry was also performed following the procedure in \cite{WerykBrown2013}. The narrow field photometry was combined with the wide field using the common time base and an offset chosen such that the narrow field matched the wide field brightness where the two overlap, producing a calibrated light curve down to a limiting magnitude of $+7^{\mathrm{M}}$.

The trajectory solution and orbit calculation followed the methodology outlined in \citet{vida2020a}, with errors found using a Monte Carlo approach. The resulting solutions show transverse residuals for both sites of just over 1~m for each event, with good agreement between stations in point-to-point speeds (cf., Appendix \ref{sec:physical_properties}, Figure \ref{fig:residuals}). Table  \ref{tab:camo_trajectories} summarizes the parameters of these solutions and the corresponding orbits. 

The resulting photometric and astrometric measurements together with the brightness and length of the wake per frame, measured as described in \citet{Stokan2013}, were then used as observational constraints for ablation modelling. The erosion model of \citet{Borovicka2007} was employed where the main fragment ablates through erosion of constituent grains. In this approach, we fix the grain density to 3000~kg~m/$^3$ and assume a constant $\Gamma$A = 1.21. The lightcurve, dynamics (velocity and deceleration), and the observed wake are then fit to the model through trial and error by varying the mass, bulk density, erosion coefficient, erosion start height, ablation coefficient, and grain mass distribution. An updated model of luminous efficiency for faint meteors which provides an empirical fit as a function of mass and speed was employed \citep{VidaOrionids}. The resulting model fits are shown graphically compared to observed data in Appendix \ref{sec:physical_properties}, Figure \ref{fig:H_mag}. Table \ref{tab:camo_results} summarizes the inferred properties of each meteoroid from this modelling. The application of the model was critical to correctly invert the initial velocity of the meteoroids - the model velocity was $\sim0.5$~km/s higher than the directly observed velocity, indicating significant deceleration occurred before the meteoroids were first observed \citep{vida2018modelling}.\\[-0.9cm]\mbox{ }

\subsection{Activity} \label{subsec:activity} 

In 2022, multiple dedicated networks collected observations of the $\tau$-Herculids meteor shower. The GMN cameras continuously recorded the shower between May 28 (solar longitude SL of 66.4$\degree$) and June 1 (SL 70.6$\degree$), except for a few hours every day due to a lack of cameras in Pacific ocean longitudes. Most visual observer data were collected on the night of May 30/31 (SL$\sim$69.17--70$\degree$), with only two additional reports of minor TAH activity around SL 67.4$\degree$ and 68.3$\degree$. The Japanese IPRMO project\footnote{\color{blue!75!yellow}{https://www.meteornews.net/2022/06/05/a-meteor-outburst-of-the-\%CF\%84-herculids-2022by-worldwide-radio-meteor-observations/}} collected radio observations of the shower between SL 67.5$\degree$ and 71.0$\degree$, and is the only continuous source of information of the TAH's activity between SL 68.7$\degree$ and 69.1$\degree$.

\begin{figure}
	\centering
	\includegraphics[width=.49\textwidth]{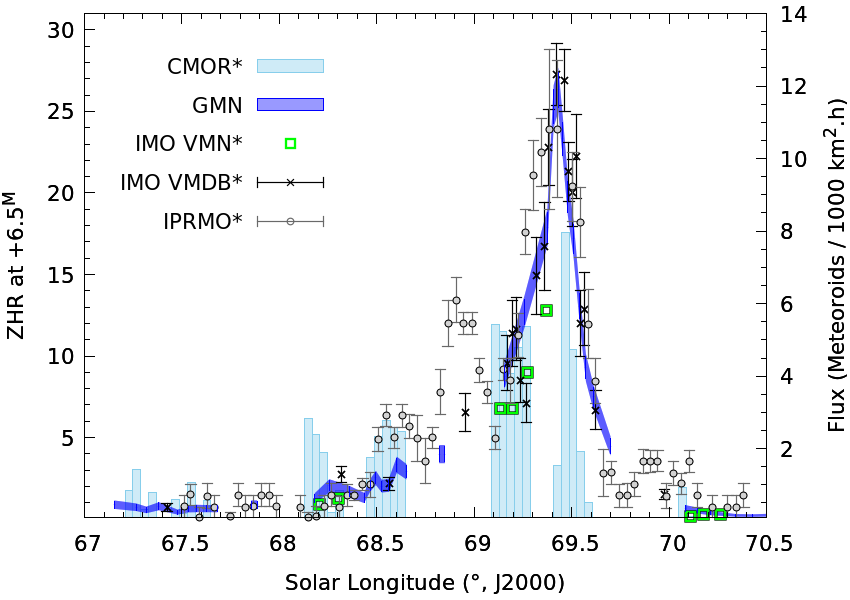}
	\caption{The activity profile of the $\tau$-Herculids in 2022, as measured by radio (CMOR, IPRMO), video (GMN, IMO VMN) and visual (IMO VMDB) networks. $^*$ZHR estimates were scaled to match the intensity level measured by GMN, selected to be the reference dataset for the analysis. 
	}
	\label{fig:2022_activity}
\end{figure}

Combining these different observations allows the reconstruction of a complete activity profile for the shower. However, the comparison of visual, video, and radar data is challenging; not only does each system suffer its own biases, but they are not equally sensitive to the same meteoroid masses. In addition, the shape and magnitude of the ZHR profiles may vary with the time resolution (choice of bin size) and the shower population index used for the computation. In order to analyze the general characteristics of the TAH, we first need to rescale each ZHR estimate to a reference set of measurements. 

With close to 1400 TAH meteors detected, the GMN network is a prolific source of observations. The magnitude of most meteors recorded by the network ranges from -3 to +4. A population index $r$ of 2.5, corresponding to a mass index $s$ of 2.0, was found to match well the observations. This is consistent with the estimates derived from visual observations, which varied from 2.5 (IMO VMDB) to 2.63-3.0\footnote{\url{https://www.imcce.fr/recherche/campagnes-observations/meteors/2022the}}. 

The meteoroid flux measured by GMN in 2022 is presented in Figure \ref{fig:2022_activity}. The flux was scaled to a limiting magnitude of +6.5 (i.e., a mass of $5 \times 10^{-3}$ g) as described in \cite{Vida2022b}. Time bins containing less than 30 meteors or a time-area product below $10^5$ km$^2$h were removed from the profile. Figure \ref{fig:2022_activity} compares the resulting ZHR with the profiles obtained by the IMO VMDB, VMN, and CMOR, all computed assuming a population index of 2.5. Measurements from the IPRMO network, for which no information about the population index was found, are also presented for comparison. 

Despite good agreement of the profiles' shape, we see some divergence between the activity levels reported by each network. The visual meteor rates reported by the IMO (VMDB and VMN), reaching a maximum of about 50 meteors per hour, were found to be 1.4 times higher than the ZHR measured by GMN. The activity levels reported by the IPRMO matched the visual observations well; however, this match is not surprising since their ZHR computation involved scaling the annual sporadic background detected with the radio instruments to visual observations\footnote{\color{blue!75!yellow}{https://www.meteornews.net/2017/07/29/the-new-method-of-estimating-zhr-using-radio-meteor-observations/}}. In contrast, the average flux measured by CMOR, which sees down to smaller sizes, exceeded the visual rates by a factor of 29. The mismatch in flux may have been exacerbated by the slow entry speed of only $\sim 15$~km~h$^{-1}$. The luminous and ionization efficiency at such low speeds change rapidly, greatly influencing any magnitude to mass conversion procedures \citep{WerykBrown2013}.

Such discrepancy in the average levels recorded by different systems is not uncommon, and has been reported for several meteor showers \citep[e.g.,][]{Egal2020}. 
In this work, we select the GMN observations to be the reference dataset for our TAH analysis. This choice is motivated by the large numbers of meteors collected by the network, the observation timespan, the resolution, and the accessibility of the data. In Figure \ref{fig:2022_activity}, we thus scaled the IMO VMDB ($\times$0.7), VMN ($\times$0.7), IPRMO ($\times$0.7) and CMOR ($\times$0.025) profiles to match the GMN flux. 

In 2022, the TAH displayed enhanced meteor activity for two days between May 29 and May 31 (SL 68.2-70.1$\degree$). Though the first TAH meteoroids may have been observed as early as May 28 (SL 66.7$\degree$), the small number of meteors recorded makes the shower hard to distinguish from the sporadic background until SL $\sim$68$\degree$. 
 
 The main peak (ZHR$\sim$27) occurred around 69.42$\pm$0.01$\degree$ (4h-4h30 UT) on May 31, and is present in visual, video, and radio data.  A secondary peak of activity was identified in IPRMO data between 15h and 19h UT on May 30 (SL 68.9-69$\degree$), reaching a ZHR of about 13. Due to the lack of observations available during this time frame (corresponding to daytime in Europe and North America), we found no confirmation of the first TAH peak in visual or video data. 
 
\cite{Ogawa2022} also highlighted the presence of a second sub-peak of activity in IPRMO data, noticeable around SL 68.549$\degree$ on May 30 (6h30 UT). The rates measured by IPRMO match well the observations performed by CMOR between SL 68.45$\degree$ and 68.65$\degree$, but we see no trace of this peak in GMN data. The low meteor rates recorded raise the possibility that this feature is simply due to observational uncertainty.
  
\subsection{Radiants}

The TAH radiants measured by CMOR and GMN in 2022 are presented in Figure \ref{fig:radiants_2022}. The figure shows the radiant distribution in geocentric and ecliptic sun-centered  coordinates, color-coded as a function of solar longitude for the GMN dataset. We observe a significant drift of the shower's radiant over the period of activity, reaching a few degrees per day. The evolution of the apparent ($\alpha,\delta$) and ecliptic ($\lambda-\lambda_\odot,\beta$) coordinates of the radiant with the solar longitude $L$, centred on the shower's maximum activity time at 69.4$\degree$, can be modelled with the following equations: 

\begin{figure}
	\centering
	\includegraphics[width=.49\textwidth]{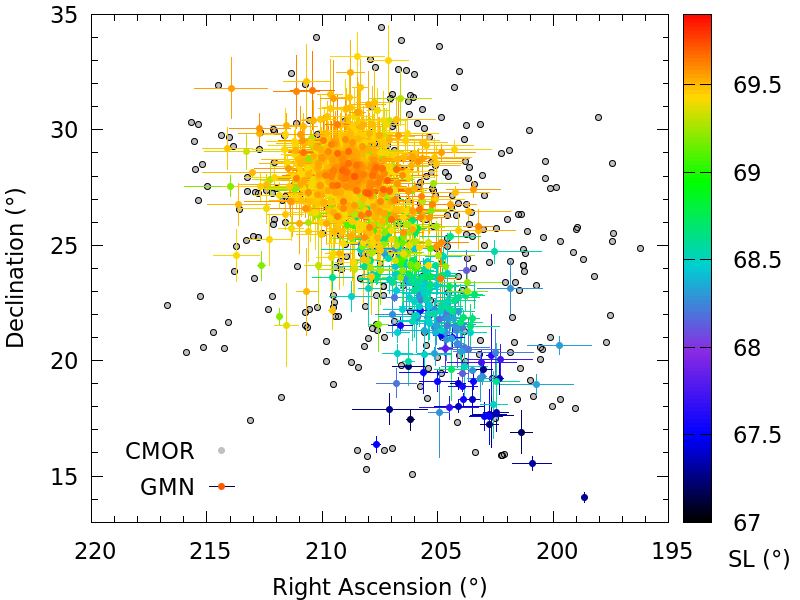}
	\includegraphics[width=.49\textwidth]{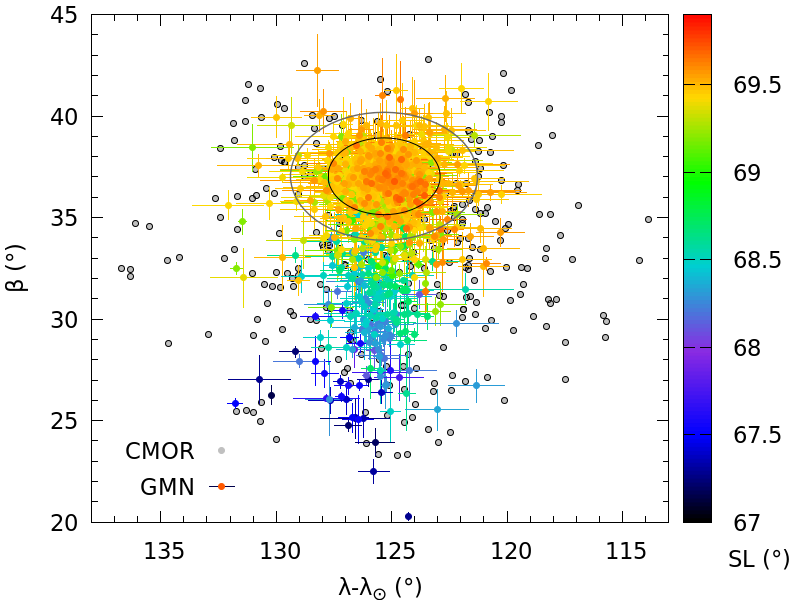}
	\caption{Geocentric (top) and ecliptic sun-centered radiants (bottom) of the TAH meteors recorded by the CMOR and GMN networks in 2022. Meteors detected by GMN are color-coded as a function of solar longitude. In the bottom plot, black and grey ellipses illustrate the 3$\sigma$ and 5$\sigma$ limits of the radiants distribution during the shower's main peak of activity (SL 69.2-69.8$\degree$), centered on ($\lambda-\lambda_\odot$,$\beta$)=(125.3$\degree$,37.0$\degree$).}
	\label{fig:radiants_2022}
\end{figure}

\begin{equation} \label{eq:radiant_drift}
\left\{\hspace{0.1cm} \begin{split}
 L & = SL-69.4\degree\\
 \alpha & = 2.83 \times L +208.64\\
 \delta & = 4.86\times L +27.68 \\
 \lambda-\lambda_\odot & = -0.68\times L + 125.34 \\
 \beta & = 5.49\times L + 36.51 \\   
 V_g & =  0.33\times L+11.36
\end{split} \right. 
\end{equation}

At the time of the second peak of activity (SL $\sim$69.2-69.8$\degree$) the ecliptic radiants in GMN data are clustered around the coordinates $\lambda-\lambda_\odot$=125.3$\degree$ and $\beta$=37.0$\degree$, with a standard deviation of 0.81$\degree$ and 0.63$\degree$ respectively. The bottom panel in Figure \ref{fig:radiants_2022} illustrates the extent of the radiant distribution at 3$\sigma$ (black ellipse) and 5$\sigma$ (grey ellipse) during the TAH main peak of activity. The dispersion of the shower \citep[median offset from the mean radiant;][]{moorhead2021meteor} was 1.2$^{\circ}$. These results are in good agreement with CAMS observations \citep{Jenniskens2022}.

\section{Model}

\subsection{Description}

The simulation of 73P's meteoroid streams follows the methodology described in \cite{Egal2019}. From a given ephemeris of the comet, thousands of particles are ejected at each apparition of 73P. The simulated meteoroids are integrated forward in time, and the distribution of the particles around Earth's orbit is examined. Meteoroids approaching Earth within a fixed distance ($DX$) and time ($DT$) criteria are retained as potential impactors. 

After selection, each particle is assigned a weight, representing the number of meteoroids that would have been released by the comet under similar ejection circumstances. The distribution of weighted impactors is used to determine the simulated shower's flux $\mathcal{F}$, which is transformed into a ZHR using the relation of \cite{Koschack1990}: 
 \begin{equation}
 \textrm{ZHR}=\frac{\mathcal{A_\text{s} F}}{(13.1r-16.5)(r-1.3)^{0.748}}
 \end{equation}
\noindent where $\mathcal{A_\text{s}}$ is the typical surface area for meteor detection by a visual observer in the atmosphere at ablation altitudes ($\mathcal{A_\text{s}}\sim 37200\; \mathrm{km}^2$) and $r$ is the measured differential population index (here fixed to 2.5). 

Finally, the timing, activity profile, and radiant distribution of the simulated shower are compared with meteor observations, in order to refine or validate the simulation's parameters selected. 

\subsection{Simulation parameters}

\subsubsection{Nucleus and ephemeris}

As in previous works \citep[e.g.,][]{Egal2019,Egal2020b}, we first examine the orbital stability of comet 73P to set the time frame of the numerical integrations. Using as our starting conditions an orbital solution provided by the Jet Propulsion Laboratory (JPL K222/7), we created one thousand clones of the comet's nominal orbit with the covariance matrix of the JPL solution. Each clone's motion was then integrated back to 1500 CE, and their orbital dispersion was examined to assess the reliability of the comet's ephemeris. 

The time evolution of the swarm of clones created for 73P is presented in Appendix \ref{app:traceability}, Figure \ref{fig:traceability}. The orbital dispersion of the  clones, highlighted by sudden increases in the swarm's standard deviation, indicates that the ephemeris of 73P prior to 1800-1820  is highly uncertain. Similar analyses conducted for different orbital solutions (e.g., before and after the 1995 break-up, with or without cometary non-gravitational forces) led to the same conclusion. In this work, we therefore restrict our numerical integrations to the period 1800 - 2050. 

In order to reduce the uncertainty on 73P's nominal evolution since 1800, (which are due to the splitting of the comet and the variable non-gravitational forces acting on the nucleus), we integrated the comet's motion using all the orbital solutions provided by the JPL for the 1930 (SAO/1930), 1979 (J7910/16), 1995 (J954/19), 1996 (K012/14), 2005 (K113/2) and 2017 (K223/8) apparitions \citep[cf.,][]{Egal2019}. After the nucleus' fragmentation in 1995, our model ephemeris describes the orbital motion of fragment 73P-C, which is assumed to be the principal remnant of the original nucleus. 

The diameter of 73P before the 1995 break-up is not accurately known; while \cite{Boehnhardt1999} estimated an upper limit of 1.1~km for the nucleus' radius, \cite{Sanzovo2001} suggested an effective radius up to 1.7~km for fragment C alone. In this work, we assume a constant radius of 1.1~km for the comet, and a bulk density of 250~kg/m$^3$ determined from CAMO measurements.  

\subsubsection{Stream formation}

Because of the comet's fragmentation history, we performed two distinct simulation sets. In the first, hereafter named ``All trails'' scenario, a new trail of meteoroids is released from the nucleus at each apparition of 73P since 1800 (and from fragment 73P-C after 1995). Meteoroids are ejected with a time step of one day for heliocentric distances below 3 AU, using the model of \cite{Crifo1997} (hereafter called CR97 model). 

For this scenario, about 528,000 particles were ejected from the comet between 1800 and 2050. Particles were equally divided among the following three size/mass/magnitude bins:
\begin{enumerate}
    \item $[10^{-4},10^{-3}]$ m, $[10^{-9},10^{-6}]$ kg, $[+15,+8]$ mag,
    \item $[10^{-3},10^{-2}]$ m, $[10^{-6},10^{-3}]$ kg, $[+8,+2]$ mag,
    \item $[10^{-2},10^{-1}]$ m, $[10^{-3},1]$ kg, $[+2,-5]$ mag.
\end{enumerate}

Our second simulation set investigates the contribution of the 1995 break-up on TAH activity. Following an approach similar to \cite{Ye2022}, we examined the orbital evolution of meteoroids ejected from the original nucleus and its main fragments at different speeds. Using as a reference the ejection speeds generated by the CR97 model (assuming a fraction of active area of 0.2), we built 6 additional simulation sets multiplying these velocities by a factor $k_{95}$ of 2, 2.5, 3.5, 4, 5 and 6.5\footnote{Following the idea of J. Vaubaillon, these ejection velocities were obtained by increasing the initial $f_a$ parameter from 0.2 to 1, 2, 3, 4, 5 and 10 respectively. }. The velocity distribution obtained for each $k_{95}$ value is presented in Appendix \ref{app:fa}, Figure \ref{fig:fa_V}. 

For each $k_{95}$ model, we ejected 15,000 meteoroids from 73P's nucleus starting at the approximate onset of the fragmentation on September 12, 1995 until December 12, 1996 (when the heliocentric distance of the model nucleus exceeds 3 AU). As previously, the meteoroids were generated with sizes comprised between 0.1-1~mm, 1-10~mm, and 1-10~cm. A bulk density of 250~kg/m$^3$ was assumed. A summary of the parameters selected for both simulation sets is provided in Table \ref{tab:simulation_parameters}.

\begin{table}
    \centering
    \begin{tabular}{ccccc}
     Sim & Radius & Density & Albedo & $f_a$  \\ \hline \hline
     All trails & 1.1 km &  250 kg/m$^3$ & 0.04 & 0.2  \\
     1995 ejecta & 1.1 km &  250 kg/m$^3$ & 0.04 & Variable  \\[0.2cm]
    \end{tabular}
    
    \begin{tabular}{ccccc}
     Sim  & N$_p$/app & N$_\text{app}$ & r$_h$ & Ejection model  \\ \hline \hline
     All trails  & 12$\times 10^3$ & 44 & $\leq$ 3 AU & [CR97] \\
     1995 ejecta  & 15$\times 10^3$ & 1 & $\leq$ 3 AU & $k_{95} \times$[CR97] \\
    \end{tabular}
    \caption{Physical characteristics of 73P's nucleus (radius, density, albedo, fraction of active area $f_a$) and meteoroids ejection parameters considered for the simulations. N$_p$/app indicates the number of meteoroids ejected at each return of the comet, $N_\text{app}$ the number of apparitions processed. Meteoroids were ejected from the nucleus within the limiting heliocentric distance $r_h$, with speeds following the model of CR97 \citep{Crifo1997} or $k_{95}$ times these velocities. }
    \label{tab:simulation_parameters}
\end{table}

\subsection{Calibration}

To derive meaningful ZHR estimates from a model, the computation of a realistic simulated flux of particles at Earth is required. As described in \cite{Egal2020b}, this is accomplished by weighting each particle as a function of the initial number of meteoroids ejected by 73P at a given epoch (and with a given size), the comet dust production over its orbit, and the meteoroid differential size frequency distribution at ejection. Our weighting scheme therefore includes several tunable parameters whose best values are determined by directly calibrating the simulated activity profiles on observations. 

For the TAH, three parameters were found to have a large influence on the simulated characteristics of the shower: the criteria used to select meteor-producing particles (distance threshold $DX$ and time threshold $DT$) and the value of the meteoroid size distribution index $u$. A careful determination of these parameters is necessary to produce reliable predictions of the shower's activity. However, the reliability of our calibration depends on the amount and quality of observations available for the shower.

Reports of TAH observations before 2022 are sparse to non-existent. Though no activity from the shower was recorded by the Harvard Radio Meteor Project in 1961-65 and 1968-69 \citep{Wiegert2005}, a few meteoroids captured on photographic plates between 1963 and 1971 were identified as possible TAH members \citep{SH1963,Lindblad1971}. Very minor TAH activity was reported by CAMS on June 2, 2011 and again on May 30-31, 2017 \citep{Rao2021}. Detectable visual activity from the shower has been reported on a single occasion by \cite{Nakamura1930}, who observed a TAH outburst of 59 meteors per hour on June 9, 1930 (SL $\sim$ 78.9$\degree$) and another event of about 72 meteors per hour the next night \citep{Jenniskens1995}. However, the unfavourable observing conditions during these nights (presence of clouds and bright moonlight) raises suspicion about the credibility of the reported intensity \citep{Wiegert2005,Rao2021}.  

Due to the paucity of TAH observations prior to 2022, we focused our modelling efforts on reproducing the characteristics of the shower in 2022, for which we have consistent records from multiple sources (cf. Section \ref{sec:observations_2022}). 

\section{Modelled 2022 activity} 
  
\subsection{Nodal-crossing locations}
   
\begin{figure}
  	\centering
  	\includegraphics[width=.5\textwidth]{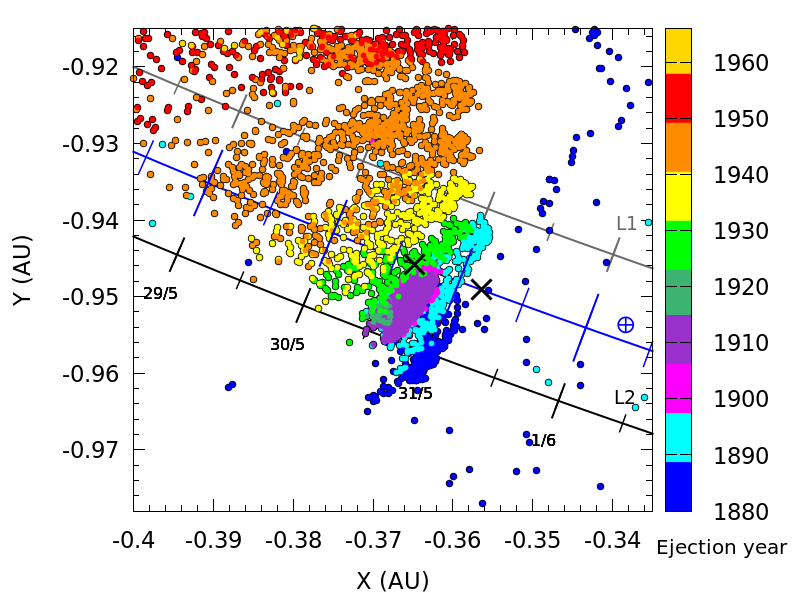}\\
  	\caption{Nodal-crossing position of simulated meteoroids reaching Earth in 2022, color-coded as a function of the ejection year. Only meteoroids located within $DT=10$ days from their node when the Earth is the closest to that position are presented. In this model, we ignore any additional meteoroid ejecta caused by 73P's break-up in 1995 (``All trails'' scenario). The black crosses along Earth's orbit (represented in blue) mark the times of the first and second TAH activity peak observed in 2022. The light and dark grey lines indicate the location of the Sun-Earth L1 and L2 Lagrange points.  }
  	\label{fig:2022_all_trails}
  \end{figure}

We first examined the influence of meteoroid trails ejected prior to the comet break-up in 1995 on the shower activity in 2022 (``All trails'' scenario). The nodal-crossing location of the simulated meteoroids, color-coded as a function of their ejection epoch, is presented in Figure \ref{fig:2022_all_trails}. Only particles crossing the ecliptic plane within $\pm$10 days of the Earth's passage are shown. The location of the Earth during the first and second TAH activity peak is indicated with black crosses in the figure. 

Within this model, only trails ejected prior to 1960 approached the Earth at the time of the meteor shower. Trails ejected by the comet between 1940 and 1948 are responsible for the slight meteor activity recorded by GMN cameras before 68.8$\degree$ SL, on May 29. In contrast, older trails may have contributed to the shower's first activity peak on May 30; in particular, the nodal footprint of the trail ejected in 1930 intersects the Earth orbit around the peak's reported time (at 69.02$\degree$ SL). However, we note that none of the trails ejected from 73P in this model can explain the TAH activity on May 31 and June 1. As a consequence, this model fails to reproduce the total duration of the shower and the main activity peak observed on May 31 (69.42$\degree$ SL).

\begin{figure}
  	\centering
  	\includegraphics[width=.5\textwidth]{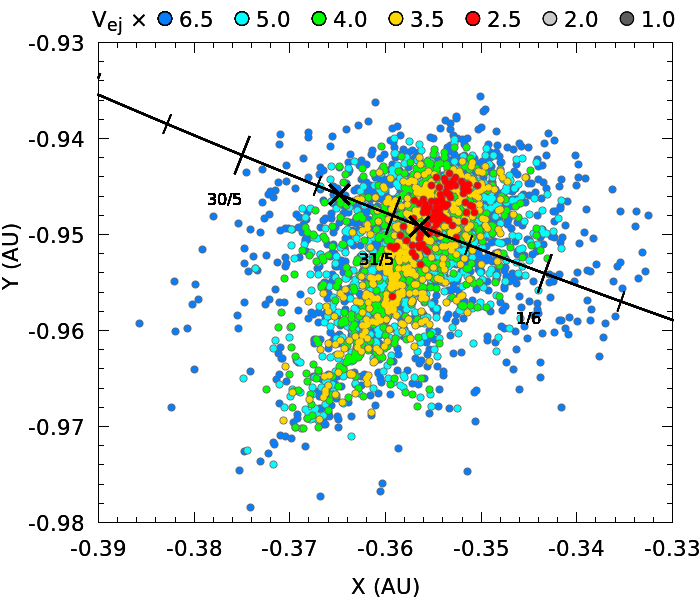}\\
  	\caption{Nodal-crossing position of the meteoroids released during the comet break-up in 1995 for different ejection speed ($V_{ej}$) multiples of the \cite{Crifo1997} model. Only meteoroids crossing the node within $DX=0.1$ AU and $DT=20$ days of the Earth's passage are presented in the plot. With these selection criteria, only meteoroids ejected with at least 2.5 the typical \cite{Crifo1997} model approached the Earth's orbit in 2022. Particles released with smaller ejection speeds (black and grey dots) did not cross the ecliptic plane in May and June in our model, and are thus not visible in the plot. }
  	\label{fig:2022_model95}
\end{figure}

Since our usual sublimation-driven meteoroid ejection model cannot by itself explain the TAH characteristics in 2022, we investigated the influence of the 1995 break-up on shower activity. Following previous authors \citep[e.g.,][]{Ye2022}, we found that meteoroids released by the fragmenting nucleus with speeds similar to gas-drag velocities have their nodes located too far from Earth to produce any meteor activity in 2022.
In Appendix \ref{app:fa}, we see that most of these particles crossed the ecliptic plane more than 0.015 AU from the Earth's orbit in 2022. In addition, none of the few particles approaching the orbit within 0.01 AU in Figure \ref{fig:fa_V} were found to cross the ecliptic planet in May or June. In our simulations, only meteoroids ejected with at least 2.5 times our reference velocity (i.e., the speed predicted by the CR97 model) were able to intersect the Earth at the time of the TAH in 2022.

Figure \ref{fig:2022_model95} presents the nodal-crossing locations in 2022 of meteoroids ejected from 73P's nucleus in 1995 for $k_{95}=\{1,2,2.5,3.5,4,5,6.5\}$ times CR97 speeds. Only particles crossing their nodes within 0.1 AU and 20 days of the Earth's passage are shown. We see that any model with $k_{95}\geq2.5$ can produce meteors at the time of the reported main peak of TAH activity. However, increased ejection speeds cause a higher dispersion of the meteoroids' nodes in 2022, which has a direct impact on the strength and duration of the predicted meteor shower. 
  
In this model, meteoroids released during the nucleus fragmentation with $k_{95}\geq4$ can also deliver some material to Earth during the shower's first activity peak. However, even the fastest simulated ejecta struggles to explain the early TAH activity. We thus suggest that both our ``All trails'' and ``1995 ejecta'' scenarios are necessary to explain the whole TAH apparition in 2022. We propose that meteoroids ejected during 73P's break-up in 1995 are responsible for the main peak of activity observed on May 31, while older trails produced the secondary peak observed a few hours earlier. 
  
\subsection{Activity}
    
To test this hypothesis, we compared the ZHR predicted from both scenarios with the observed activity profile in Figure \ref{fig:2022_activity}. The values of $DX$ and $DT$ required to select meteor-producing particles were determined by matching the dispersion of the simulated radiants with observations. We found that retaining meteoroids approaching Earth with $DX=0.005$ AU and $DT=10$ days in 2022 provided the best agreement with the GMN radiants (see Section \ref{subsec:modelled_radiants}), without drastically reducing the number of selected particles.
  
\begin{figure}[!ht]
  	\centering
  	\includegraphics[width=.49\textwidth]{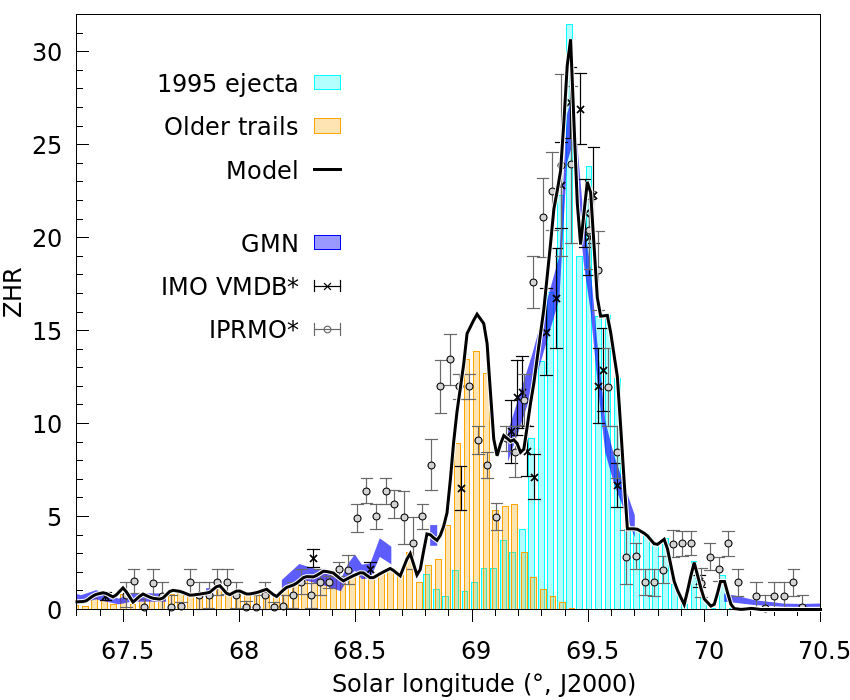}
  	\caption{Modelled activity of the 2022 TAH, considering all the meteoroids ejected from 73P since 1800 (orange profile) and the material released during the comet's break-up in 1995 with $k_{95}$=4 (blue profile). The combined profile (black line) is compared with measurements performed by visual (IMO VMDB), radio (IPRMO) and video (GMN) networks.}
  	\label{fig:2022_model_profile}
\end{figure}

In total, about 20,200 particles were retained for the ZHR computation in 2022. The profile obtained when considering all the meteoroids ejected from the comet since 1800 is presented in Figure \ref{fig:2022_model_profile} (orange boxes). 
  
After analysis of the different $k_{95}$ outputs, we noticed that all the $k_{95}\geq2.5$ models (processed with the same weighting scheme) predicted similar activity variations and dates of maximum meteor rates. As indicated from the nodal distribution in Figure \ref{fig:2022_model95}, the main difference between the $k_{95}$ models is in regards to the shower's predicted duration. In our simulations, we find that the stream ejected from the fragmenting nucleus with $k_{95}$=4 is most consistent with the observations; the corresponding profile is illustrated with blue boxes in Figure \ref{fig:2022_model_profile}. 
      
The complete simulated profile, determined from both the ``All trails'' and ``1995 ejecta'' data and represented by the black line in Figure \ref{fig:2022_model_profile}, is in excellent agreement with the observed activity. In particular, we find that meteoroids ejected during the comet break-up accurately reproduce the characteristics reported for the main peak, including its time, shape, radiant location and duration.

The predicted time of the first peak ($SL\sim69.02\degree$), caused by material released between 1900 and 1947, is shifted by about 2.5 hours compared with the IPRMO measurements. The shape of the simulated profile between SL 68.7$\degree$ and 69.1$\degree$ also somewhat diverges from the observations, and displays no enhanced activity around 68.549$\degree$. We note that these discrepancies mainly relate to trails that were ejected prior to the comet discovery, while the activity due to younger trails (e.g., 1941 \& 1947) is consistent with the rates reported. This may hint at inaccuracy of the ephemeris used for the comet prior to 1930. However, given the small ZHR and the lack of additional observations during this time frame, we consider that our model still satisfactorily explains the shower's first peak of activity. 
 
The best estimate of the meteoroids' size distribution index at ejection is $u=3.9$ for the old trails and $u=3.7$ for the 1995 ejecta. It is encouraging to note that the latter value is consistent with Spitzer observations of the comet \citep{Vaubaillon2010}, and with the model developed by \cite{Ye2022}. 
  
\subsection{Radiants} \label{subsec:modelled_radiants}
  
\begin{figure}[!ht]
  	\centering
  	\includegraphics[width=.49\textwidth]{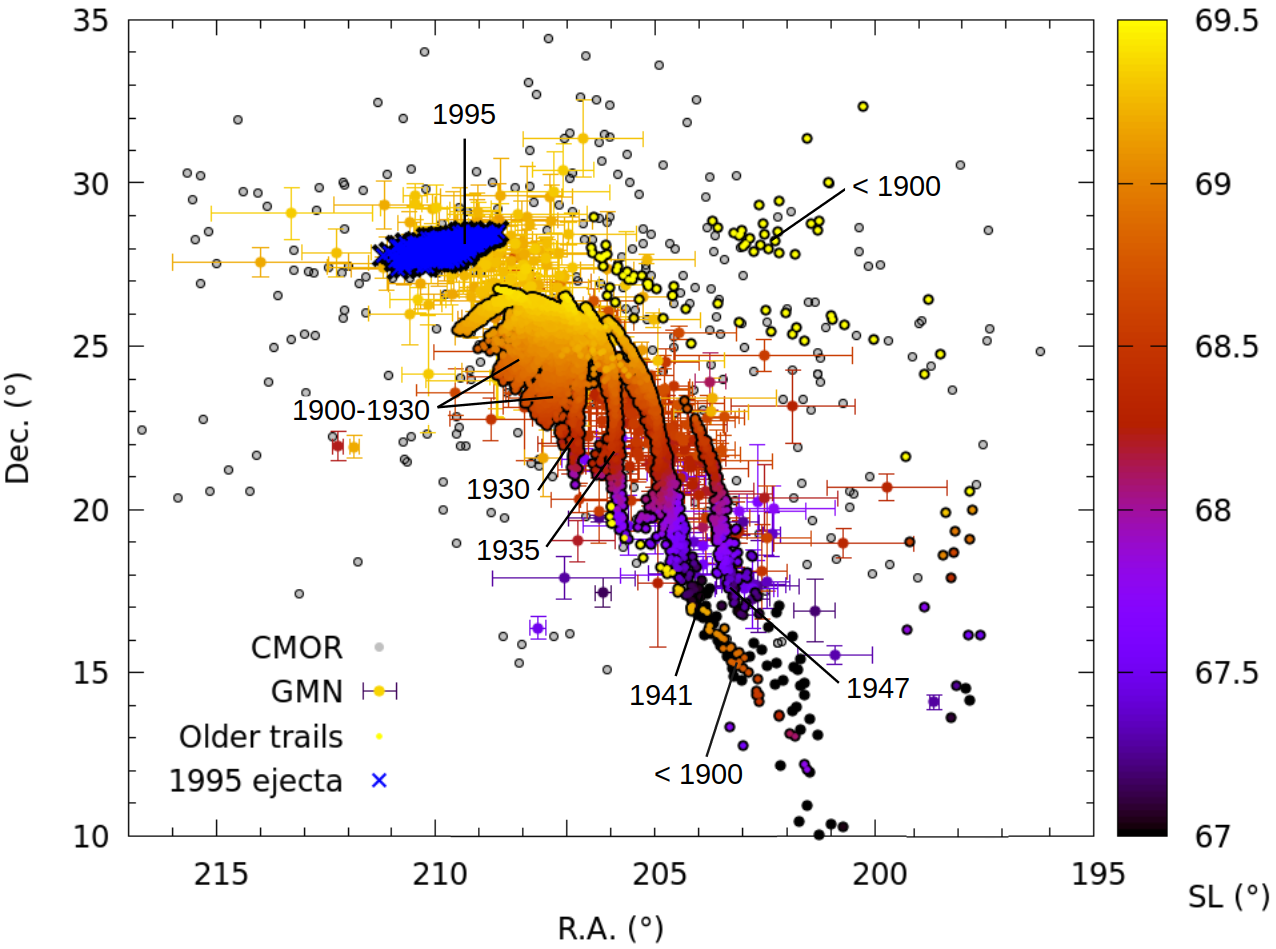}
  	\caption{Geocentric radiants of the simulated meteoroids in 2022, released at each apparition of 73P since 1800 (``All trails'' scenario, colored circles) or during the 1995 fragmentation (``1995 ejecta'' model, blue crosses). Only particles approaching the Earth within $DX=0.005$ AU and $DT=10$ days are presented. The simulated radiants are compared with the observations performed by GMN and CMOR presented in Figure \ref{fig:radiants_2022}.}
  	\label{fig:2022_model_radiants}
\end{figure}

The simulated radiant distribution is presented in Figure \ref{fig:2022_model_radiants}, along with GMN and CMOR observations. Each trail ejected from the comet prior to 1950 produces an elongated radiant structure, that is less diffuse than the observed radiants. However, the location, timing, and overall dispersion of the simulated radiants match well the GMN observations for SL$\in[67\degree,69.3\degree]$. 
      
The radiants produced by the 1995 ejecta ($k_{95}=4$ model) are indicated with blue crosses in the figure. Once again, the radiants' location is consistent with the values measured by GMN during the TAH second activity peak ($SL\sim69.4\degree$), but the particles' dispersion is smaller than the observed one. Though simulations performed at larger ejection speeds (e.g., $k_{95}=6.5$) were found to slightly increase the radiant dispersion in right ascension, none of our $k_{95}$ models precisely reproduce the spread measured by the GMN. 

The Earth's gravity, by bending the trajectory of the meteoroids, directly influences the scattering of the apparent radiants observed. This effect is more pronounced for long meteors, that enter the atmosphere at shallow angles or with low velocities like the TAH. The curvature of the meteoroids' trajectory due to gravity, ignored in our simulations, can partially explain the discrepancy between the observed and modelled distributions of the apparent radiants around SL 69.4$\degree$.

However, the gaps in the modelled radiant distribution imply that our simulations do not include all the meteoroids detected on Earth in 2022. We thus explored alternative scenarios for stream formation, involving different ephemeris solutions for the comet or the ejection of additional material from other fragments of the original nucleus (fragments 73P-A, 73P-B, 73P-C and 73P-E). However, none of these simulations led to a better agreement with the observations, and were not further investigated. 

The sun-centered ecliptic radiants of the simulated stream in 2022 is presented in Appendix \ref{app:ecliptic_radiant}, Figure \ref{fig:modelled_ecliptic_radiants}. In the figure, the particles are color-coded as function of their geocentric velocity, and compared with CMOR and GMN measurements. At the time of the main activity peak (SL$\sim$69.4$\degree$), all the meteoroids retained in our model possess a geocentric velocity of 12$\pm$0.1 km/s. This value is remarkably consistent with the velocities determined from CAMO's high-resolution measurements (cf. Table \ref{tab:camo_trajectories}). 
   
Despite the partial incompleteness of the simulated radiant distribution, our simulations are in good agreement with the shower activity, duration and radiant observed by different detection networks (cf. Figures \ref{fig:2022_model_profile}, \ref{fig:2022_model_radiants} and \ref{fig:modelled_ecliptic_radiants}). We thus conclude that the combination of material ejected during the comet break-up in 1995 with older meteoroid trails successfully explains the overall characteristics of the TAH in 2022.

\subsection{Size distribution}

\begin{figure} 
	\centering
	\includegraphics[width=.5\textwidth]{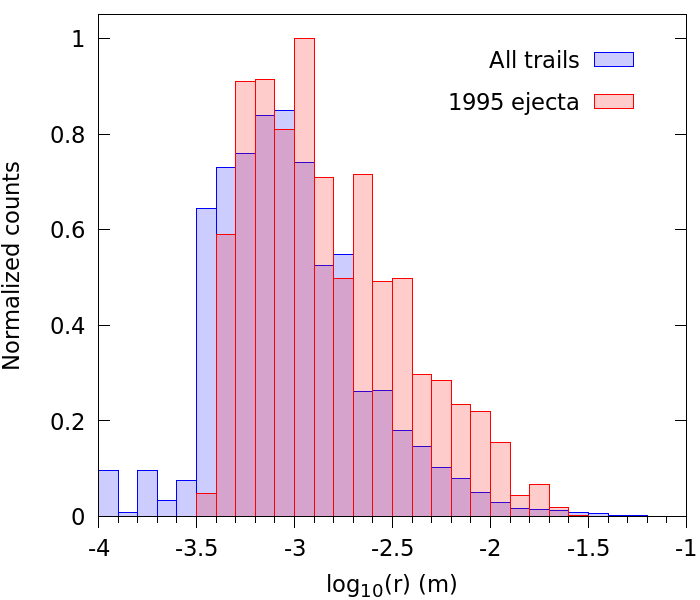}\\
	\caption{Differential size distribution of the simulated TAH in 2022. Meteoroids ejected during the comet 1995 break-up (``1995 ejecta'' model) or earlier (``All trails'' model) are represented in red and blue. While most particles reaching the Earth in 2022 are a few millimeters in size, the presence of several cm-sized particles in both distributions (-3.5 to +2 in mag) may explain the multiple detections of bright TAH meteors. }
	\label{fig:size_distrib}
\end{figure}

The size distribution of the simulated meteors in 2022 is shown in Figure \ref{fig:size_distrib}.  We see that sub-mm and mm-class meteoroids efficiently reach the Earth in both models, with most particles being a few millimeters in radius. Meteoroids ejected during the 1995 break-up may have produced meteors of magnitude +12 to -1.5, caused by particles of radius comprised between 0.3~mm and 3~cm. Older meteoroids of 0.1~mm to 6~cm in size may also have created meteors of magnitude +15 to -3.5 in 2022. In contrast with \cite{Ye2022}, we find that several cm-sized particles can reach the Earth in both models, in particular in the ``1995 ejecta'' simulation. This is consistent with the numerous detections of bright TAH meteors reported in 2022.

\section{Extension of the model}

\subsection{Postdictions} \label{sec:postdictions}

\begin{figure}[!ht]
	\centering
	\includegraphics[width=.48\textwidth]{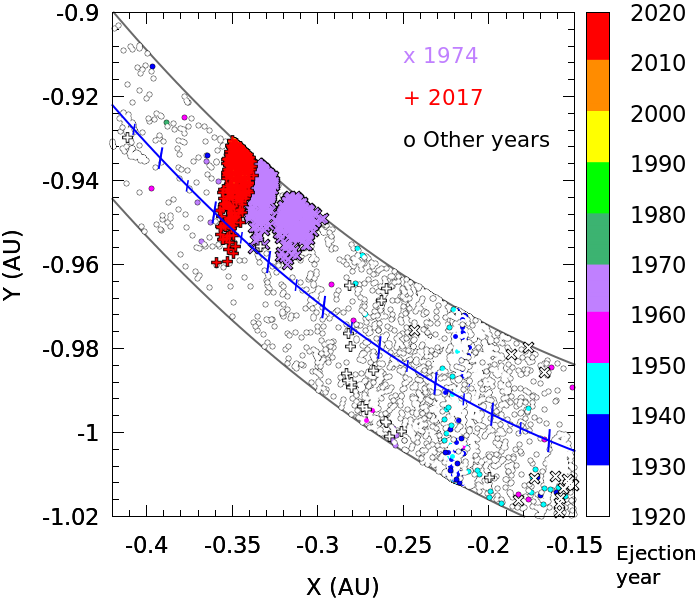}\\
	\includegraphics[width=.48\textwidth]{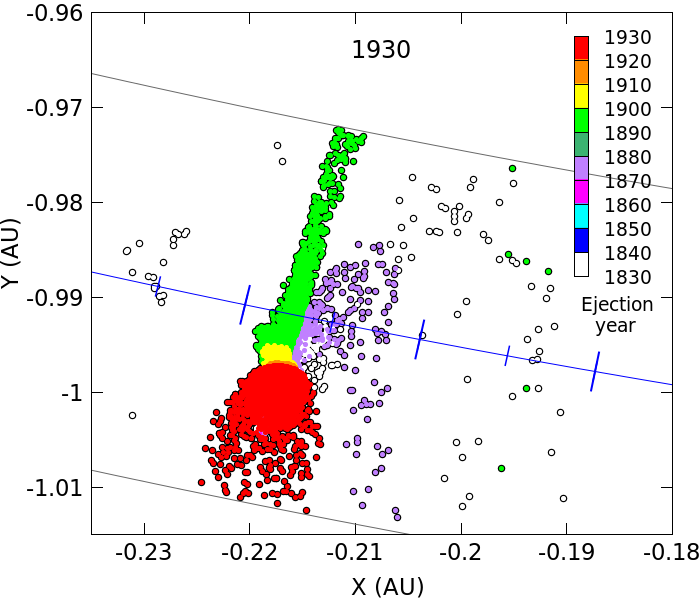}\\
	\caption{Nodal-crossing locations of the meteoroids ejected from comet 73P since 1800 and arriving between 1931 and 2021. Meteoroids approaching the Earth's orbit (blue line) within $DT=20$ days and $DX=0.02$ AU (grey lines) are presented. The nodes are color-coded as a function of the ejection year. }
	\label{fig:past_nodes}
\end{figure}

Despite the low number of available TAH observations prior to 2022, we investigated the possible past activity of the meteor shower using our  2022 calibrated model. The nodal-crossing locations of the simulated streams between 1930 and 2021 are presented in Figure \ref{fig:past_nodes}. Though significant TAH activity is absent for most of the examined years, Figure \ref{fig:past_nodes} highlights three favourable apparitions of the meteor shower in 1930, 1974, and 2017. 

In our simulations, the strongest TAH return occurred in 1930, when several trails ejected from the comet intersected the Earth's orbit. The proximity of the 1892 trail may have led to a few hours of significant meteor activity on June 8, around $77.73\degree$ SL. With the weighting scheme determined for the 2022 apparition, the model predicts maximum rates of about 300 meteors per hour (see Figure \ref{fig:1930_model} in Appendix \ref{app:predicted_profiles}).  However, any estimate of the shower's precise strength in 1930 needs to be considered with caution: because of the concentration of several different trails around the Earth's path, we found that ZHR predictions vary substantially with the value of $DX$ considered. For example, increasing $DX$ from 0.005 AU to 0.02 AU raises the predicted maximum rates by an order of magnitude in our model.

Despite this uncertainty, our simulations are in good agreement with the conclusions of \cite{Luthen2001} who predicted a possible activity from the 1892 trail around 77.75$\degree$. In the model of \cite{Wiegert2005}, most of the material also crossed the Earth's orbit one day before the outburst date reported by Nakamura, with only a few particles remaining close to the Earth around $78.9\degree$. Since no observation of the TAH was conducted on June 8, 1930 \citep{Rao2021}, the veracity of our model can not be tested for this specific day. Like previous authors, we are reduced to suggesting that Nakamura's observations on June 9 and 10 relate to the diffuse trails ejected in 1880 and prior to 1830. The possibility of an erroneous timing report, placing the activity on June 9-10 instead of June 8-9 could also reconcile the model with the observations \citep{Wiegert2005}. 

After the 1930 outburst, the model predicts no significant activity until recent years. With our selection criteria $DX$ of 0.005 AU, no detectable TAH meteors would have been observed in 1974, despite the proximity of the dense 1963-64 trail (purple crosses in Figure \ref{fig:past_nodes}). This is consistent with non observations of the shower reported in IAU Circular 2672. In contrast, the model displays minor activity in 2017 caused by the stream ejected during 73P's previous perihelion return in 2011-2012. The simulated ZHR profile of the 2017 apparition (cf., Appendix \ref{app:predicted_profiles}, Figure \ref{fig:1930_model}) confirms that a ZHR level of a few TAH per hour are postdicted to have occurred between SL 69.3$\degree$ and 70.2$\degree$ on May 30-31, which is consistent with CAMS observations.  

\subsection{Predictions}

Despite some timing uncertainties in the modelled 1930 outburst, our simulations reproduce most of the TAH characteristics (i.e., the shower's apparition years, radiant location, and intensity). This encouraged us to extend the model to the future and to forecast $\tau$-Herculid activity until 2050. Figure \ref{fig:future_nodes} in Appendix \ref{app:future_nodes} shows the nodal crossing locations of the simulated stream for future years of interest. Meteoroids ejected since 1800 are shown in the top inset and those ejected during the 1995 break-up are shown in the lower inset. Frequently, both simulation sets predict activity in the same years, including 2023, 2029, 2033, 2035, 2041, 2047, and 2049. 

\begin{figure}
  \includegraphics[width=.48\textwidth]{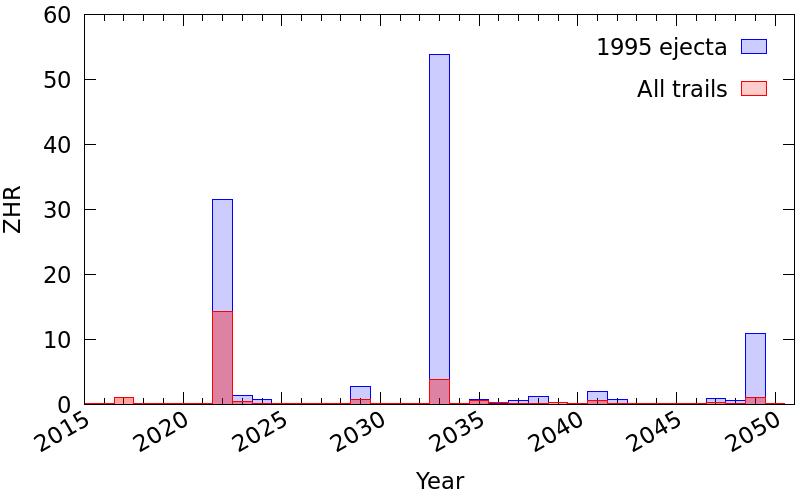}
  \caption{Predicted TAH activity between 2015 and 2050, caused by meteoroids ejected during the break-up of 73P in 1995 (blue boxes) or from other trails (red boxes).}
  \label{fig:future_activity}
\end{figure}

The predicted ZHR rates, based on the calibration performed on the 2022 apparition, are illustrated in Figure \ref{fig:future_activity}. Several minor TAH apparitions (ZHR $<$ 5) are expected in the coming years, including in 2023, 2029, 2035, 2037, or 2041. The model also predicts a significant TAH return in 2033 (ZHR$\sim$55) and moderate activity in 2049 (ZHR$\sim$10). In both cases, most of the activity is expected to originate from the comet break-up in 1995, although trails ejected between 1920 and 2010 also contribute to the shower. The predicted profiles for 2033 and 2049 are presented in Figure \ref{fig:future_profiles} in Appendix \ref{app:predicted_profiles}.

In contrast with 2022, future TAH apparitions are expected to occur earlier in the year, in late April and early May. In 2033, most of the activity is predicted to occur between 45$\degree$ and 50.5$\degree$ SL, peaking around 48$\degree$ SL on May 8. A second peak of activity may also occur around 52$\degree$ SL, caused by trails ejected between 2027 and 2029. The periods of activity are expected to last longer than in 2022 due to the  elongated shapes of the nodes in the ecliptic plane (cf. Figure \ref{fig:future_nodes}). As noted by \cite{Wiegert2005}, the integrated flux of meteoroids in 2049 may thus exceed the 2022 level, despite lower ZHR values. 

\section{Discussion}

Developing an accurate model of the $\tau$-Herculid meteoroid stream is a complex task. Reliable stream simulations rely on a good knowledge of the parent comet's position and activity in the past, sometimes over several centuries. Like many JFCs, 73P has suffered multiple close encounters with Jupiter in the past decades, making its ephemeris prior to 1800 highly uncertain. In addition, the comet has undergone a series of disintegration events since 1995, generating hundreds of fragments from the original nucleus. 

Because of the comet's erratic history, we find that a simple meteoroid stream model based on standard ice-sublimation mechanisms can not reproduce the TAH characteristics in 2022. While the first activity peak around 69.02$\degree$ SL could have been caused by trails ejected from 73P before 1960, the main peak observed at 69.42$\degree$ is best modelled with meteoroids released during the nucleus break-up in 1995. In agreement with \cite{Ye2022}, we find that meteoroids ejected from the nucleus at 4 times the velocities of \cite{Crifo1997} provide the best match to the peak's time and duration. In particular, many meteoroids ejected from the nucleus at speeds between 20 and 70 m/s intersected the Earth's orbit at the time of the TAH maximum activity. 

In contrast, we suggest that older meteoroid trails are necessary to explain the TAH characteristics prior to 69.2$\degree$, including the shower duration and the extent of the radiants distribution. However, the reliability of our model's calibration is intrinsically linked to the quantity and quality of observations available. Due to the lack of complementary records obtained with visual and video networks, the characteristics of the TAH peak observed by the IPRMO around SL 68.9$\degree$ are less certain than for the main peak.

Although the shape of the IPRMO profile matches well the GMN, CMOR and IMO VMDB results before 68.5$\degree$ and after 69.2$\degree$ (cf. Figure \ref{fig:2022_activity}), measurement errors in the data \citep{Ogawa2022} could affect the precise timing and magnitude of the first peak reported around 68.9$\degree$. As a consequence, the calibration of our ``All trail'' model may not be optimal and vary with future observations of the shower. However, since our simulations do not predict any strong activity caused by these trails in the future, we expect the accuracy of the IPRMO measurements to have only a moderate influence on our TAH forecast.

Though necessary to explain the observations, the combination of two distinct models does reduce the robustness of the model's calibration. Indeed, the weighting solution applied to the modelled stream is not unique, and different combinations of the coefficients $(u,DX,DT)$ can lead to similar results for a specific apparition of the shower. 

Determining these unknown parameters with confidence generally requires comparing the simulations' output with long-term observations of the meteor shower. However, because most TAH observations were performed in 2022, our ZHR predictions need to be considered with caution. This is well illustrated by the modelled 1930 rates, which varied significantly with the weighting scheme considered (see Section \ref{sec:postdictions}). Continued observations of the shower will help improve the reliability of future TAH models.

\section{Conclusion} \label{sec:conclusion}

This work presents detailed modelling of the $\tau$-Herculid apparition in 2022 calibrated using observed activity. The combination of observations performed in the visual, video, and radio mass ranges allows us to reconstruct a complete activity profile for the shower between May 28 and June 1. In 2022, the TAH displayed two peaks of activity. Maximum meteor rates were observed by multiple networks near 69.42$\degree$ SL on May 31, while a secondary activity peak was detected in IPRMO data around 69.02$\degree$. The analysis of two meteoroids observed with CAMO close to the maximum activity time and likely associated with the 1995 breakup, indicates that TAH meteoroids are fragile; the estimated bulk density of 250 km/m$^3$ is among the lowest values measured by the CAMO instrument so far. 

The numerical simulation of meteoroids ejected at each apparition of comet 73P since 1800 indicates that old meteoroid trails are responsible for the first TAH activity peak observed in 2022, as well as additional  apparitions of the shower in 1930 and 2017. 
In contrast, the second peak of activity in 2022 was probably caused by meteoroids ejected during the comet break-up in 1995. We find that particles released from the splitting nucleus with 4 times the typical gas-drag velocities of \cite{Crifo1997} reproduce well the shape, duration, and intensity of the second peak. Though a few model particles do approach the Earth's orbit in other years, our simulations calibrated with the 2022 activity profile postdict no significant TAH activity (ZHR$\geq$5) in the 1930-2022 period.

The extension of our model to future years predicts possible returns of the shower in 2023, 2029, 2033, 2035, 2041, and 2049. While ZHR values smaller than 5 are expected for most of these apparitions, stronger rates of 40-55 and 10 meteors per hour could be reached in 2033 and 2049, respectively. In both cases, most of the activity is expected to originate from meteoroids ejected during the 1995 break-up, with a minor contribution from trails ejected from fragment 73P/C between 1920 and 2010. Visual, video, and radar observations of future $\tau$-Herculid apparitions are strongly encouraged to provide additional constraints on numerical models.

\section{Acknowledgements} \label{sec:acknowledgments}

We are very thankful to J. Vaubaillon and Q. Ye for fruitful discussions and model cross-checks that enhanced our analysis, and to the reviewer for his comments that helped improving this manuscript. We would also like to thank the following GMN station operators whose cameras provided the data used in this work and contributors who made important code contributions (in alphabetical order): Richard Abraham, Victor Acciari, Rob Agar, Yohsuke Akamatsu, David Akerman, Daknam Al-Ahmadi, Jamie Allen, Alexandre Alves, Don Anderson, Željko Andreić, Martyn Andrews, Enrique Arce, Georges Attard, Chris Baddiley, David Bailey, Erwin van Ballegoij, Roger Banks, Hamish Barker, Jean-Philippe Barrilliot, Ricky Bassom, Richard Bassom, Alan Beech, Dennis Behan, Ehud Behar, Josip Belas, Alex Bell, Florent Benoit, Serge Bergeron, Denis Bergeron, Jorge Bermúdez Augusto Acosta, Steve Berry, Adrian Bigland, Chris Blake, Arie Blumenzweig, Ventsislav Bodakov, Robin Boivin, Claude Boivin, Bruno Bonicontro, Fabricio Borges, Ubiratan Borges, Dorian Božičević, David Brash, Stuart Brett, Ed Breuer, Martin Breukers, John W. Briggs, Gareth Brown, Peter G. Brown, Laurent Brunetto, Tim Burgess, Jon Bursey, Yong-Ik Byun, Ludger Börgerding, Sylvain Cadieux, Peter Campbell-Burns, Andrew Campbell-Laing, Pablo Canedo, Seppe Canonaco, Jose Carballada, Steve Carter, David Castledine, Gilton Cavallini, Brian Chapman, Jason Charles, Matt Cheselka, Enrique Chávez Garcilazo, Tim Claydon, Trevor Clifton, Manel Colldecarrera, Michael Cook, Bill Cooke, Christopher Coomber, Brendan Cooney, Jamie Cooper, Andrew Cooper, Edward Cooper, Rob de Corday Long, Paul Cox, Llewellyn Cupido, Christopher Curtis, Ivica Ćiković, Dino Čaljkušić, Chris Dakin, Fernando Dall'Igna, James Davenport, Richard Davis, Steve Dearden, Christophe Demeautis, Bart Dessoy, Pat Devine, Miguel Diaz Angel, Paul Dickinson, Ivo Dijan, Pieter Dijkema, Tammo Dijkema Jan, Luciano Diniz Miguel, Marcelo Domingues, Stacey Downton, Stewart Doyle, Zoran Dragić, Iain Drea, Igor Duchaj, Jean-Paul Dumoulin, Garry Dymond, Jürgen Dörr, Robin Earl, Howard Edin, Raoul van Eijndhoven, Ollie Eisman, Carl Elkins, Ian Enting Graham, Peter Eschman, Nigel Evans, Bob Evans, Bev M. Ewen-Smith, Seraphin Feller, Eduardo Fernandez Del Peloso, Andres Fernandez, Andrew Fiamingo, Barry Findley, Rick Fischer, Richard Fleet, Jim Fordice, Kyle Francis, Jean Francois Larouche, Patrick Franks, Stefan Frei, Gustav Frisholm, Jose Galindo Lopez, Pierre Gamache, Mark Gatehouse, Ivan Gašparić, Chris George, Megan Gialluca, Kevin Gibbs-Wragge, Marc Gilart Corretgé, Jason Gill, Philip Gladstone, Uwe Glässner, Chuck Goldsmith, Hugo González, Nikola Gotovac, Neil Graham, Pete Graham, Colin Graham, Sam Green, Bob Greschke, Daniel J. Grinkevich, Larry Groom, Dominique Guiot, Tioga Gulon, Margareta Gumilar, Peter Gural, Nikolay Gusev, Kees Habraken, Alex Haislip, John Hale, Peter Hallett, Graeme Hanigan, Erwin Harkink, Ed Harman, Marián Harnádek, Ryan Harper, David Hatton, Tim Havens, Mark Haworth, Paul Haworth, Richard Hayler, Andrew Heath, Sam Hemmelgarn, Rick Hewett, Nicholas Hill, Lee Hill, Don Hladiuk, Alex Hodge, Simon Holbeche, Jeff Holmes, Steve Homer, Matthew Howarth, Nick Howarth, Jeff Huddle, Bob Hufnagel, Roslina Hussain, Jan Hykel, Russell Jackson, Jean-Marie Jacquart, Jost Jahn, Nick James, Phil James, Ron James Jr, Rick James, Ilya Jankowsky, Alex Jeffery, Klaas Jobse, Richard Johnston, Dave Jones, Fernando Jordan, Romulo Jose, Edison José Felipe Pérezgómez Álvarez, Vladimir Jovanović, Alfredo Júnior Dal’Ava, Javor Kac, Richard Kacerek, Milan Kalina, Jonathon Kambulow, Steve Kaufman, Paul Kavanagh, Ioannis Kedros, Jürgen Ketterer, Alex Kichev, Harri Kiiskinen, Jean-Baptiste Kikwaya, Sebastian Klier, Dan Klinglesmith, John Kmetz, Zoran Knez, Korado Korlević, Stanislav Korotkiy, Danko Kočiš, Bela Kralj Szomi, Josip Krpan, Zbigniew Krzeminski, Patrik Kukić, Reinhard Kühn, Remi Lacasse, Gaétan Laflamme, Steve Lamb, Hervé Lamy, Jean Larouche Francois, Ian Lauwerys, Peter Lee, Hartmut Leiting, David Leurquin, Gareth Lloyd, Robert Longbottom, Eric Lopez, Paul Ludick, Muhammad Luqmanul Hakim Muharam, Pete Lynch, Frank Lyter, Guy Létourneau, Angélica López Olmos, Anton Macan, Jonathan Mackey, John Maclean, Igor Macuka, Nawaz Mahomed, Simon Maidment, Mirjana Malarić, Nedeljko Mandić, Alain Marin, Bob Marshall, Colin Marshall, Gavin Martin, José Martin Luis, Andrei Marukhno, José María García, Keith Maslin, Nicola Masseroni, Bob Massey, Jacques Masson, Damir Matković, Filip Matković, Dougal Matthews, Phillip Maximilian Grammerstorf Wilhelm, Michael Mazur, Sergio Mazzi, Stuart McAndrew, Lorna McCalman, Alex McConahay, Charlie McCormack, Mason McCormack, Robert McCoy, Vincent McDermott, Tommy McEwan, Mark McIntyre, Peter McKellar, Peter Meadows, Edgar Mendes Merizio, Aleksandar Merlak, Filip Mezak, Pierre-Michael Micaletti, Greg Michael, Matej Mihelčić, Simon Minnican, Wullie Mitchell, Georgi Momchilov, Dean Moore, Nelson Moreira, Kevin Morgan, Roger Morin, Nick Moskovitz, Daniela Mourão Cardozo, Dave Mowbray, Andrew Moyle, Gene Mroz, Brian Murphy, Carl Mustoe, Juan Muñoz Luis, Przemek Nagański, Jean-Louis Naudin, Damjan Nemarnik, Attila Nemes, Dave Newbury, Colin Nichols, Nick Norman, Philip Norton, Zoran Novak, Gareth Oakey, Perth Observatory Volunteer Group, Washington Oliveira, Jorge Oliveira, Jamie Olver, Christine Ord, Nigel Owen, Michael O’Connell, Dylan O’Donnell, Thiago Paes, Carl Panter, Neil Papworth, Filip Parag, Ian Parker, Gary Parker, Simon Parsons, Ian Pass, Igor Pavletić, Lovro Pavletić, Richard Payne, Pierre-Yves Pechart, Holger Pedersen, William Perkin, Enrico Pettarin, Alan Pevec, Mark Phillips, Anthony Pitt, Patrick Poitevin, Tim Polfliet, Renato Poltronieri, Pierre de Ponthière, Derek Poulton, Janusz Powazki, Aled Powell, Alex Pratt, Miguel Preciado, Nick Primavesi, Paul Prouse, Paul Pugh, Chuck Pullen, Terry Pundiak, Lev Pustil’Nik, Dan Pye, Nick Quinn, Chris Ramsay, David Rankin, Steve Rau, Dustin Rego, Chris Reichelt, Danijel Reponj, Fernando Requena, Maciej Reszelsk, Ewan Richardson, Martin Richmond-Hardy, Mark Robbins, Martin Robinson, David Robinson, Heriton Rocha, Herve Roche, Paul Roggemans, Adriana Roggemans, Alex Roig, David Rollinson, Andre Rousseau, Jim Rowe, Nicholas Ruffier, Nick Russel, Dmitrii Rychkov, Robert Saint-Jean, Michel Saint-Laurent, Clive Sanders, Jason Sanders, Ivan Sardelić, Rob Saunders, John Savage, Lawrence Saville, Vasilii Savtchenko, Philippe Schaak, William Schauff, Ansgar Schmidt, Yfore Scott, James Scott, Geoff Scott, Jim Seargeant, Jay Shaffer, Steven Shanks, Mike Shaw, Jamie Shepherd, Angel Sierra, Ivo Silvestri, François Simard, Noah Simmonds, Ivica Skokić, Dave Smith, Ian A. Smith, Tracey Snelus, Germano Soru, Warley Souza, Jocimar Justino de Souza, Mark Spink, Denis St-Gelais, James Stanley, Radim Stano, Laurie Stanton, Robert D. Steele, Yuri Stepanychev, Graham Stevens, Richard Stevenson, Thomas Stevenson, Peter Stewart, William Stewart, Paul Stewart, Con Stoitsis, Andrea Storani, Andy Stott, David Strawford, Claude Surprenant, Rajko Sušanj, Damir Šegon, Marko Šegon, Jeremy Taylor, Yakov Tchenak, John Thurmond, Stanislav Tkachenko, Eric Toops, Torcuill Torrance, Steve Trone, Wenceslao Trujillo, John Tuckett, Sofia Ulrich, Jonathan Valdez Aguilar Alexis, Edson Valencia Morales, Myron Valenta, Jean Vallieres, Paraksh Vankawala, Neville Vann, Marco Verstraaten, Arie Verveer, Jochen Vollsted, Predrag Vukovic, Aden Walker, Martin Walker, Bill Wallace, John Waller, Jacques Walliang, Didier Walliang, Christian Wanlin, Tom Warner, Neil Waters, Steve Welch, Tobias Westphal, Tosh White, Alexander Wiedekind-Klein, John Wildridge, Ian Williams, Mark Williams, Guy Williamson, Graham Winstanley, Urs Wirthmueller, Bill Witte, Jeff Wood, Martin Woodward, Jonathan Wyatt, Anton Yanishevskiy, Penko Yordanov, Stephane Zanoni, Pető Zsolt, Dario Zubović and Marcelo Zurita.  This work was supported in part by NASA Meteoroid Environment Office under cooperative agreement 80NSSC21M0073 and by the Natural Sciences and Engineering Research Council of Canada (Grants no. RGPIN-2016-04433 \& RGPIN-2018-05659), and by the Canada Research Chairs Program.


\bibliography{References}{}
\bibliographystyle{aasjournal}


\appendix

\section{Physical properties}\label{sec:physical_properties}

 \begin{figure}[!ht]
  	\centering
  	\includegraphics[width=.5\textwidth, angle=90]{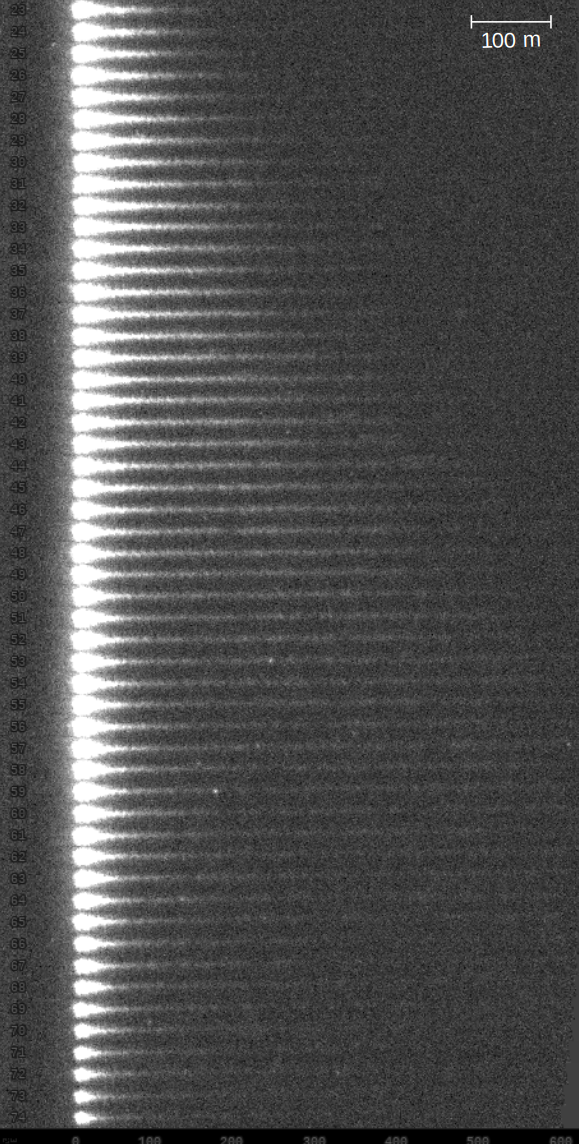}
  	\caption{Video stack of the CAMO tracked $\tau$-Herculid occurring at 033542 UT on May 31, 2022. Shown in each column are the aligned meteor and its wake. The time progresses from left to right at 10~ms increments. The left column begins at a height of 87.3~km and each subsequent picture is approximately 150~m lower in height. The final image on the right is at an altitude of 79.7~km. Scale on the y-axis is in m - the full length from bottom to top in the image is 700~m at the range to the meteor. Here a noticeable wake can be followed for almost 500~m in many frames, a clear indication of significant fragmentation.}
  	\label{fig:tau_herc}
  \end{figure}
  
\begin{table}[!ht]
    \centering
    \begin{tabular}{cccccccccc}
    \hline
    \hline
     Event & $\alpha_g$ ($\degree$)  & $\delta_g$ ($\degree$) &  $v_g$ (km/s) & q (AU) & a (AU) & e & i ($\degree$) & $\Omega$ ($\degree$) & $\omega$ ($\degree$) \\[0.1cm]
    \hline
     20220531\_032732 & 209.330 & 28.265 & 12.033 & 0.990 & 3.081 & 0.679 & 11.110 & 69.410 & 199.781 \\
      & $\pm$ 0.007 & $\pm$ 0.018 & $\pm$ 0.005 & $\pm$ 0.000 & $\pm$ 0.003 & $\pm$ 0.000 & $\pm$ 0.007 & $\pm$ 0.000 & $\pm$ 0.012  \\[0.1cm]
     20220531\_033552 & 209.410 & 28.537 & 12.007 & 0.990 & 3.050 & 0.675 & 11.161 & 69.415 & 199.702\\
     & $\pm$ 0.026 & $\pm$ 0.071 & $\pm$ 0.018 & $\pm$ 0.000 & $\pm$ 0.009 & $\pm$ 0.001 & $\pm$ 0.027 & $\pm$ 0.000 & $\pm$ 0.043 \\[0.1cm]
    \hline
    \hline
    \end{tabular}
    \caption{Geocentric radiant ($\alpha_g$,$\delta_g$) and velocity ($v_g$) of two $\tau$-Herculid meteors observed with CAMO. The heliocentric orbital elements (q, a, e, i, $\Omega$, $\omega$) of the meteoroids in J2000 are also presented. The stated uncertainties correspond to the 1-$\sigma$ formal uncertainties of the fit.  }
    \label{tab:camo_trajectories}
\end{table}

\begin{figure}[!ht]
    \centering
    \includegraphics[width=\textwidth]{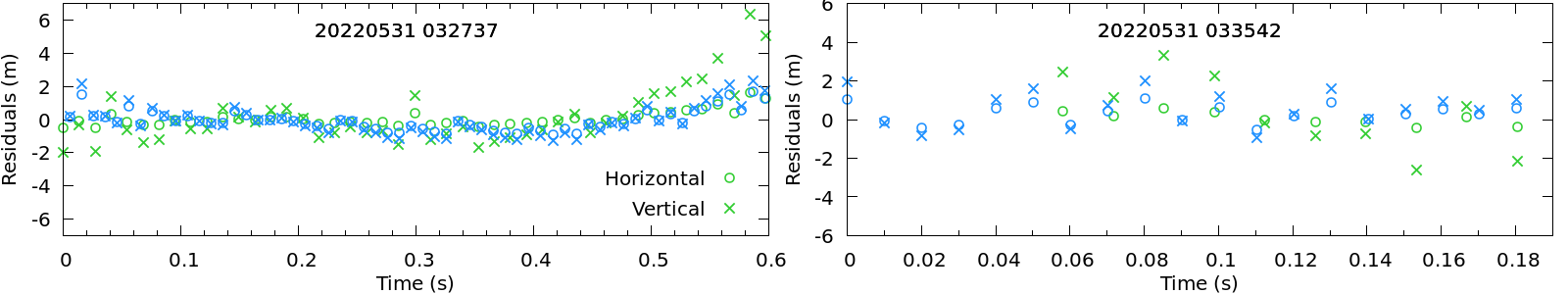}
    \caption{Trajectory fit residuals (horizontal and vertical) of the events 20220531\_032732 \& 20220531\_033552, computed for station 1 (blue) and station 2 (green) with respect to a straight line aligned with the radiant  provided in Table \ref{tab:camo_trajectories}.}
    \label{fig:residuals}
\end{figure}
  
  \newpage

\begin{figure}[!ht]
    \centering
    \includegraphics[trim={0 0 0 0.78cm}, clip, height=9.5cm]{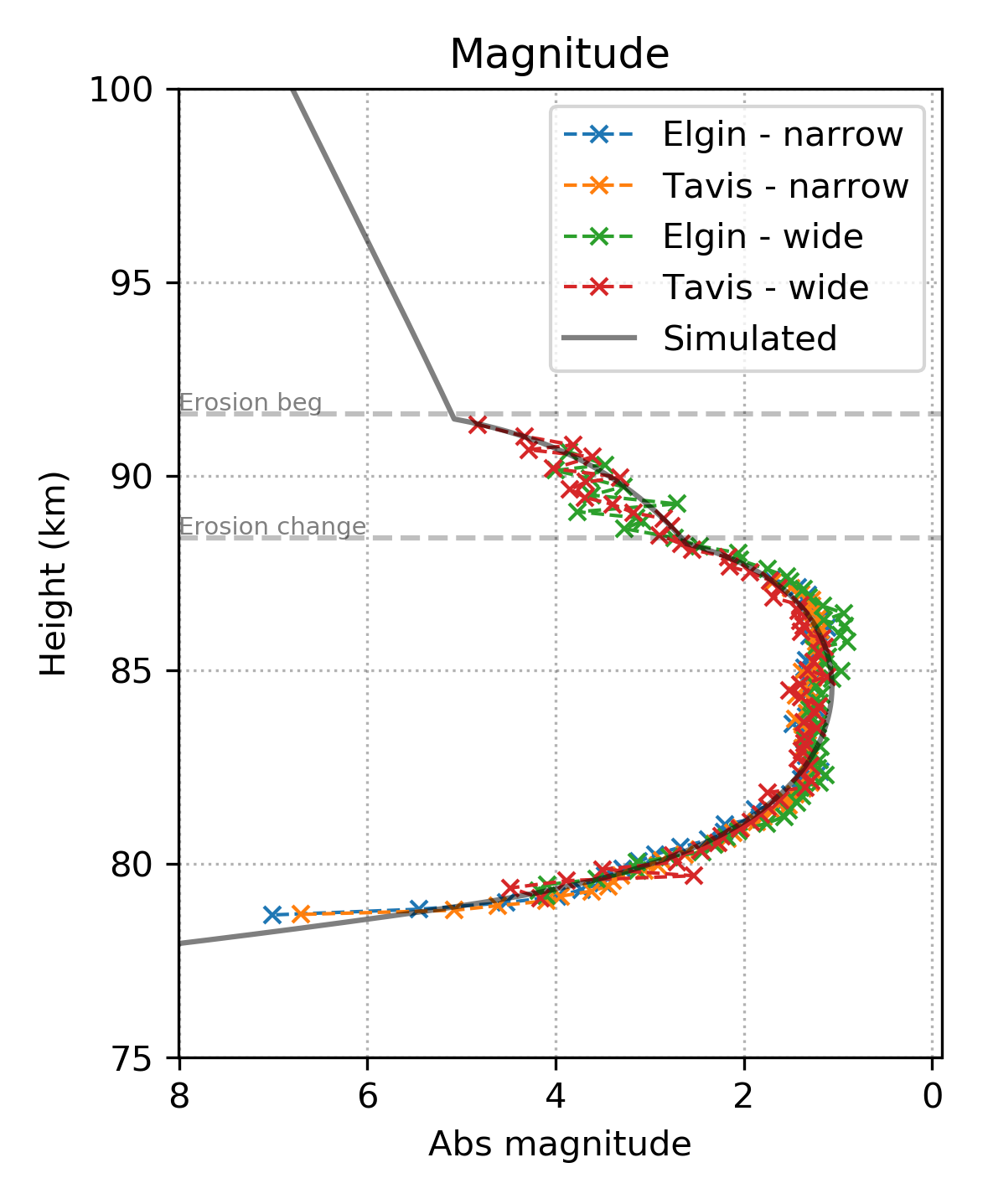}
    \includegraphics[trim={0.9cm 0 0 0.78cm}, clip, height=9.5cm]{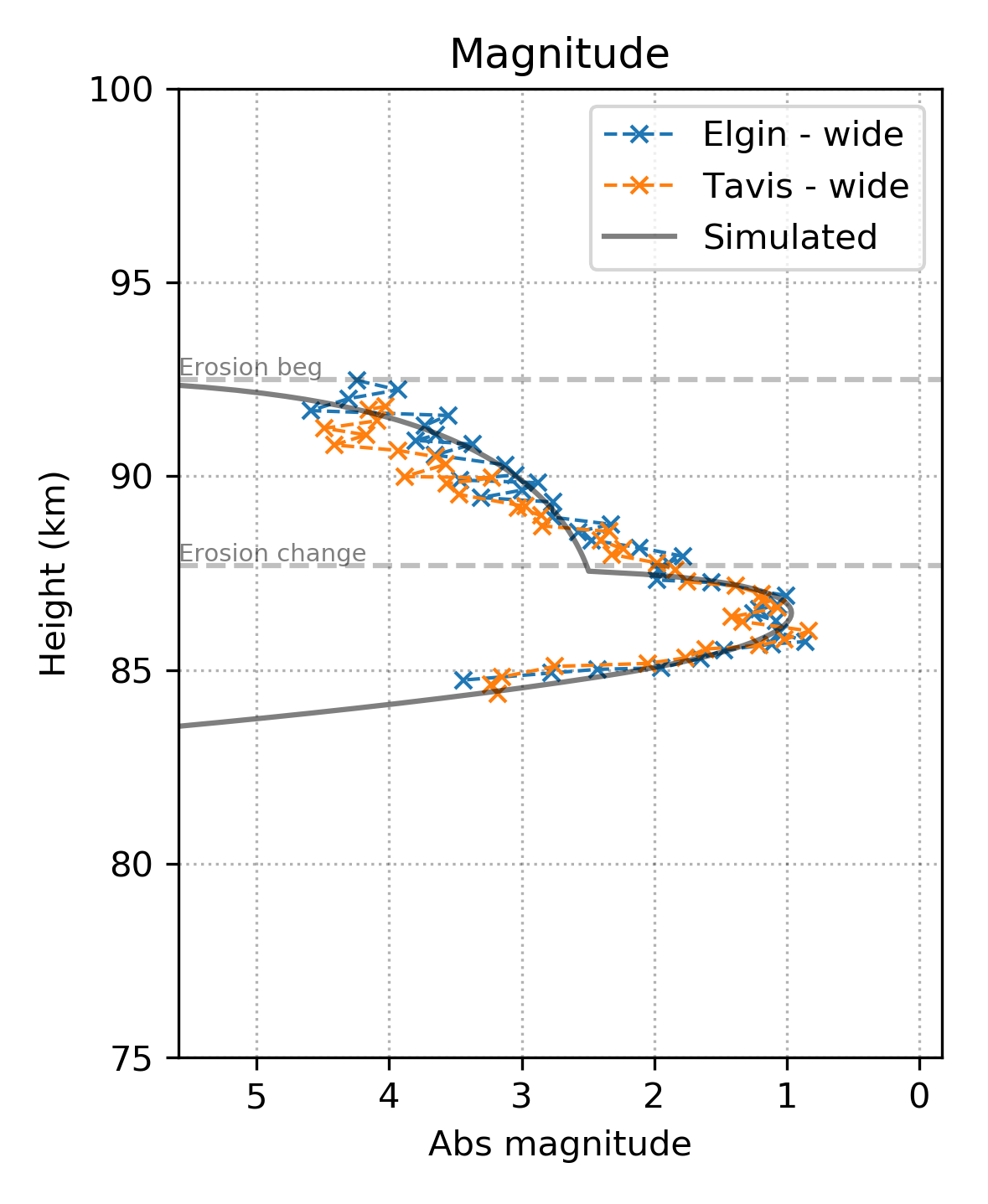}\\
    \caption{Observed and simulated light curve of the two CAMO events (left: 20220531\_032732 and right: 20220531\_033552). Colored dots in the figure refer to observations performed at the two CAMO sites (Elginfield and Tavistock) with both the wide-field and the narrow-field cameras. Both events experienced a change (a second ``bump'') in the erosion coefficient at the height of $\sim$88~km, indicating the beginning of more vigorous erosion. The black line indicates the best fit model to both light curves. }
    \label{fig:H_mag}
\end{figure}

\begin{table}[!ht]
    \centering
    \begin{tabular}{ccccccccc}
    \hline
    \hline
     Event &  V$_\text{init}$  &  m$_\text{init}$ & $\rho_\text{bulk}$ & $\sigma$ & H$_\text{E1}$ & $\sigma_\text{E1}$ & H$_\text{E2}$ & $\sigma_\text{E2}$ \\[0.15cm]
      & km/s & kg & kg/m$^3$ & kg/MJ & km & kg/MJ & km & kg/MJ \\
    \hline
     20220531\_032732 & 16.34 & 1.2$\times10^{-4}$ & 230 & 0.030 & 91.6 & 0.04 & 88.40 & 0.12 \\
     20220531\_033552 & 16.31 & 5.0$\times10^{-5}$ & 250 & 0.035 & 92.5 & 0.07 & 87.70 & 0.55 \\
    \hline
    \hline
    \end{tabular}
    \caption{Modelled physical properties of the two meteoroids observed with CAMO. V$_\text{init}$ is the initial velocity at the top of the atmosphere,  m$_\text{init}$ is the initial mass, $\rho_\text{bulk}$ is the meteoroid bulk density, $\sigma$ is the ablation coefficient, H$_\text{E1}$ is the height at which erosion began with $\sigma_\text{E1}$ erosion coefficient, which changed at height H$_\text{E2}$ to $\sigma_\text{E2}$.    }
    \label{tab:camo_results}
\end{table}

\clearpage
\newpage
\section{Measured GMN flux in 2022} \label{app:flux}
\vspace{-0.2cm}
\begin{figure*}[!ht]
    \centering
    \includegraphics[width=\textwidth]{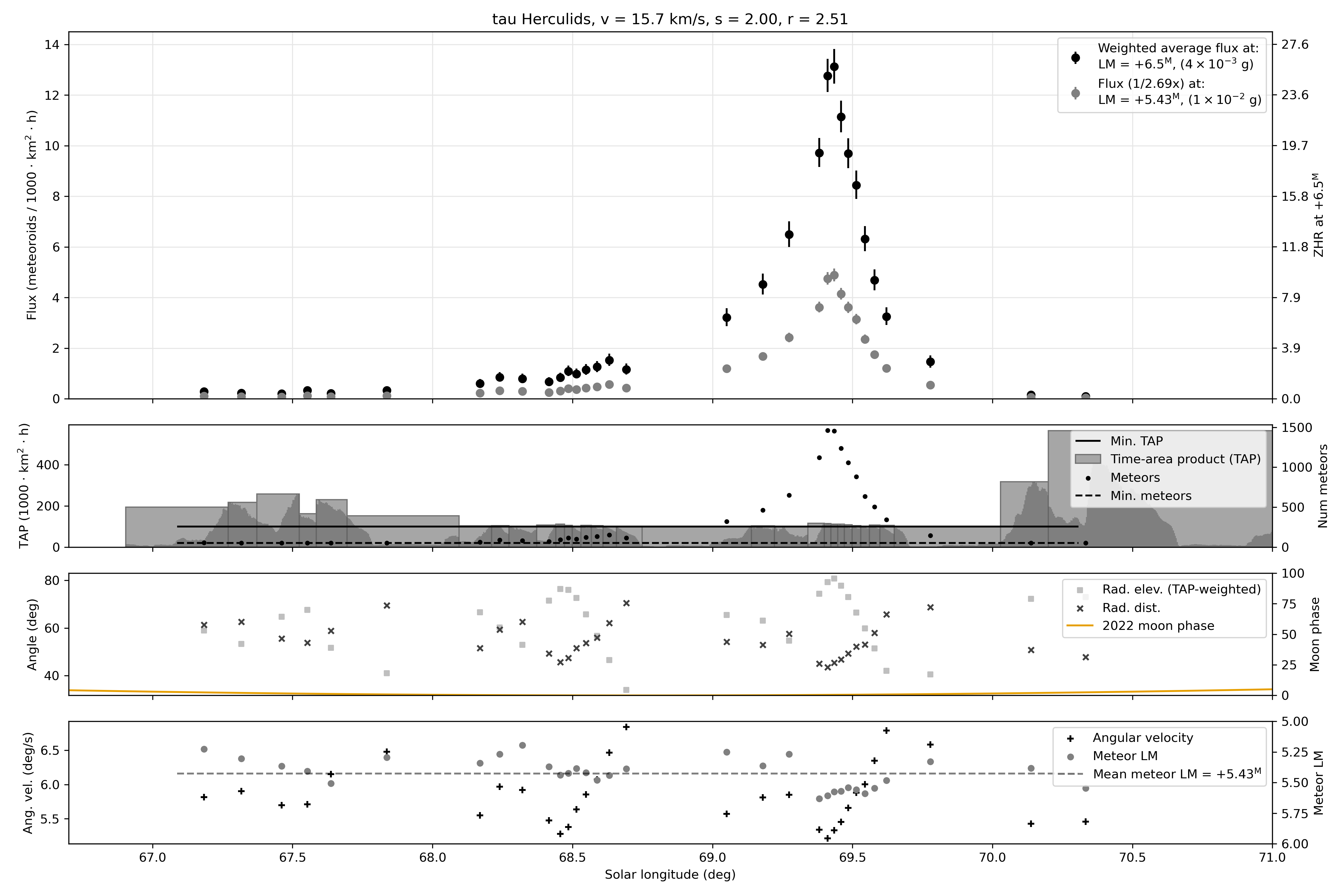}
    \caption{GMN flux measurements of the TAH outburst. The lengths of bins were determined by setting a minimum time-area product (100,000~km$^2$~h) and a minimum number of meteors (30) in each bin. Error bars for some points in the wings are smaller than plot markers. The top inset shows the flux and confidence interval for a reference mass of $4 \times 10^{-3}$ g (black dots) while the grey symbols represent the flux to a limiting mass of $1 \times 10^{-2}$ g, which was derived from the average effective meteor limiting sensitivity of the aggregated network data ($+5.43^{\mathrm{M}}$). A mass index of $s = 2.0$ was used for flux scaling. The second inset from the top shows the total available time-area product (TAP) in each bin and the TAP distribution inside each bin (dark grey histograms). The reference solar longitude of each flux measurement is weighted by the TAP inside the bin. Note that the GMN had virtually no coverage during the first peak at 69$^{\circ}$ and was thus not captured in the graph. The black dots are the total number of raw meteor detections by all cameras. The third inset from the top shows the average radiant elevation across all cameras and the radiant distance from the center of the camera field of view, both weighted by the TAP. The phase of the moon is also shown for ease of interpretation, with 100 being a full moon. Finally, the bottom inset shows the average limiting magnitude and average meteor shower angular velocity in the center of the field of view per bin, TAP weighted.}
    \label{fig:2022_flux}
\end{figure*}
\vspace{-0.2cm}

\clearpage
\newpage
\section{Traceability} \label{app:traceability}
\vspace{-0.2cm}
\begin{figure*}[!ht]
    \centering
    \includegraphics[width=.99\textwidth]{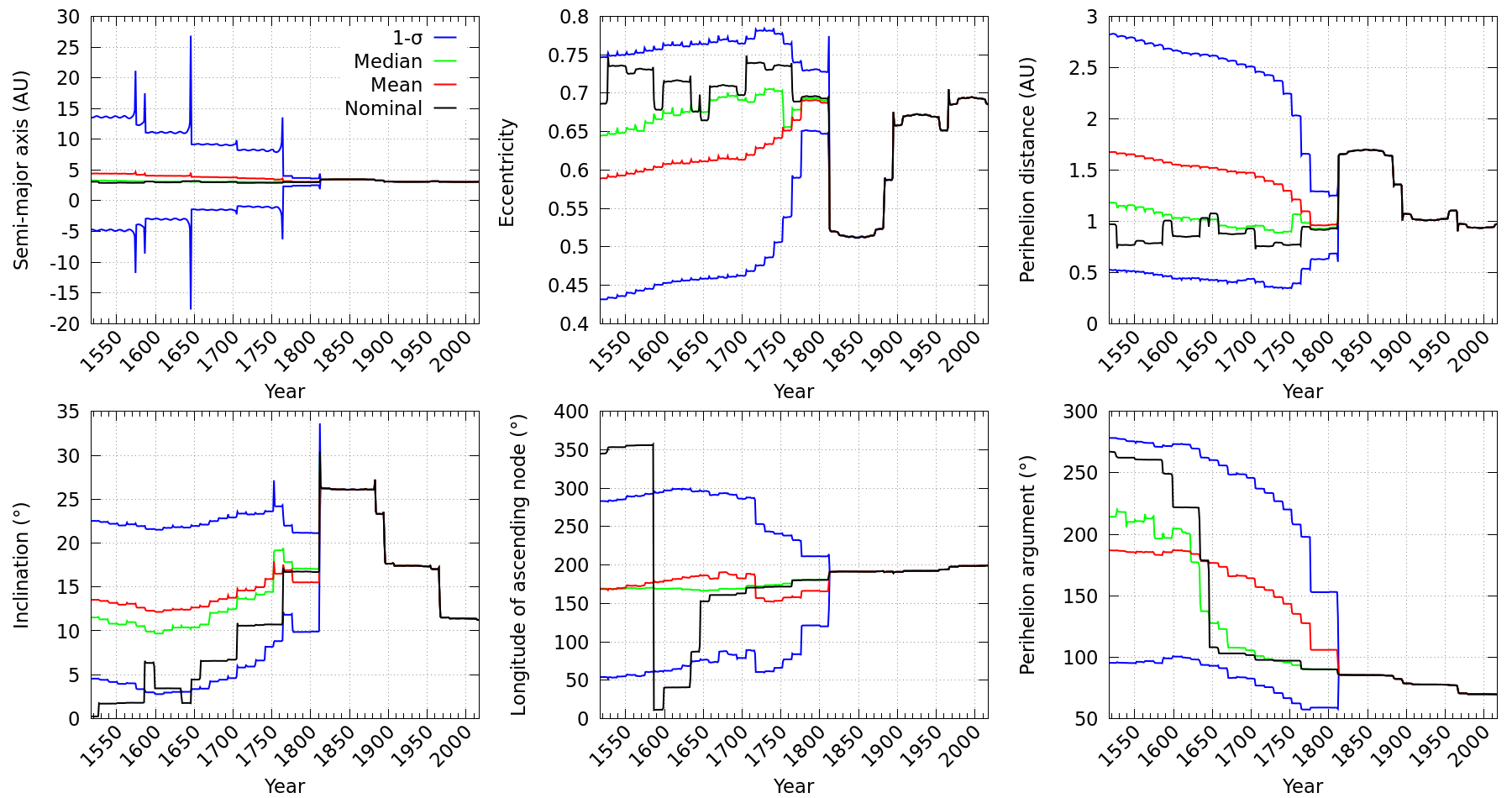}
    \caption{Example of traceability analysis for comet 73P, using the JPL K222/7 orbital solution. A thousand clones of the nominal orbit (black line) are created using the orbital covariance matrix and integrated until 1500 CE. The orbital dispersion of the swarm of clones, characterized by the standard deviation (blue lines), indicates that the ephemeris of 73P prior to 1810 is highly uncertain. }
    \label{fig:traceability}
\end{figure*}

\vspace{-0.6cm}
\section{$k_{95}$ models} \label{app:fa}
\vspace{-0.2cm}

\begin{figure*}[!ht]
	\centering
	\includegraphics[width=.49\textwidth]{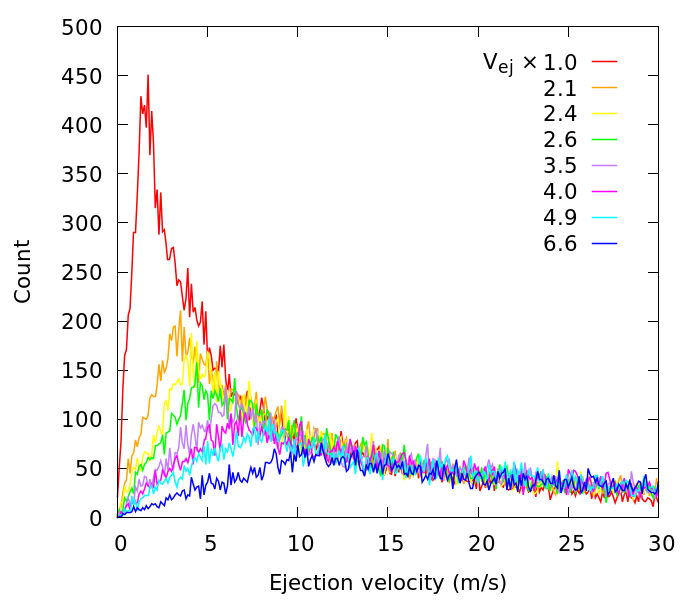}
	\includegraphics[width=.49\textwidth]{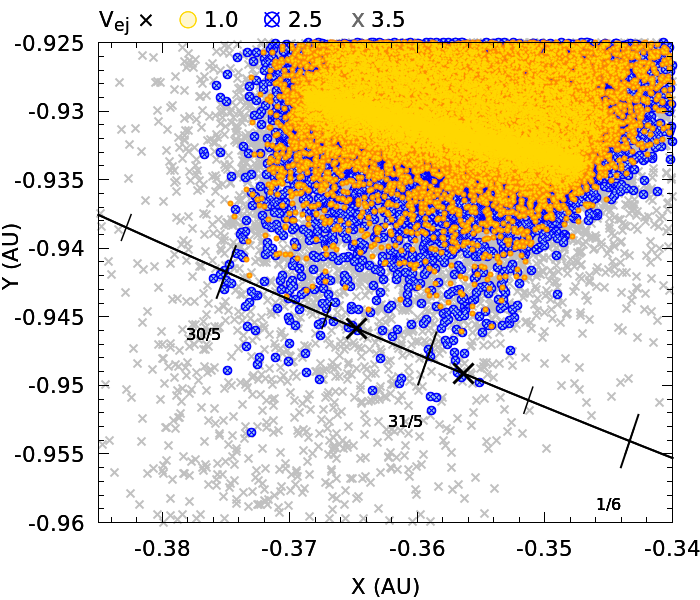}
	\caption{Left: velocity distribution of the meteoroids ejected from comet 73P with the model of  \cite{Crifo1997} (red line), or with $k_{95}=2$ to $k_{95}=6.5$ times the model ejection speeds. Right: all nodal-crossing locations in 2022 of the meteoroids released from 73P's nucleus in 1995 with velocities of $k_{95}=$1, 2.5 and 3.5 times the speeds predicted by the model of \cite{Crifo1997}. Only particles ejected with at least 2.5 times \cite{Crifo1997}'s original values are able to approach the Earth's orbit in 2022.}    
	\label{fig:fa_V}
\end{figure*}

\clearpage
\newpage

\section{Modelled radiants and velocities in 2022} \label{app:ecliptic_radiant}

\begin{figure}[!ht]
    \centering
    \includegraphics[width=.7\textwidth]{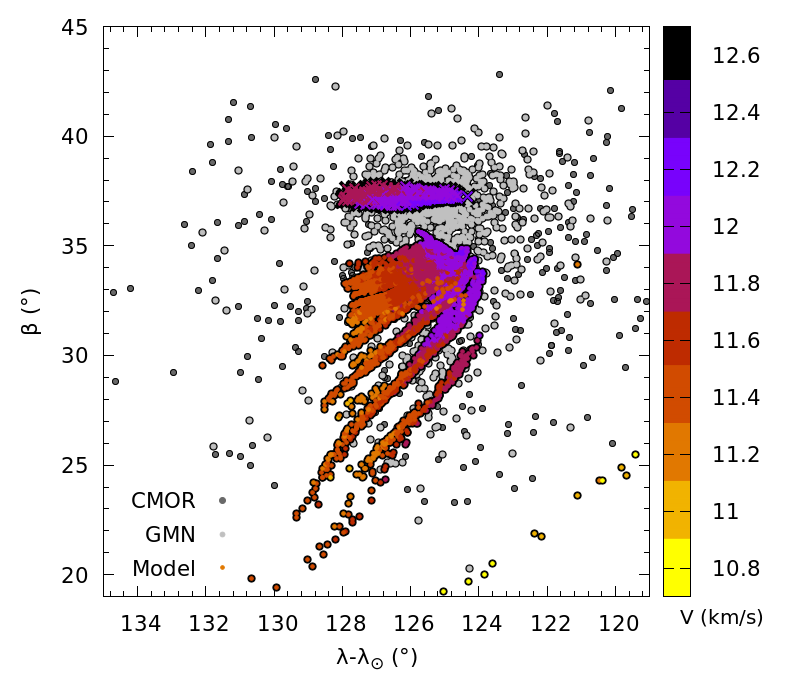}
    \caption{Sun-centered ecliptic radiants of the modelled apparition of the $\tau$-Herculids in 2022, from material ejected during 73P's break-up in 1995 or at previous apparitions of the comet. The simulated radiants are color-coded as function of the meteoroids' geocentric velocities, and compared with the GMN and CMOR data of Figure \ref{fig:radiants_2022}.  All the particles contributing to the TAH main activity peak around the coordinates ($\lambda-\lambda_\odot$,$\beta$)=(125.3$\degree$,37.0$\degree$) posses a geocentric velocity of 12$\pm$0.1 km/s, which is consistent with the velocities measured by CAMO  (cf. Table \ref{tab:camo_trajectories}).}
    \label{fig:modelled_ecliptic_radiants}
\end{figure}

\clearpage
\newpage

\section{Future nodal-crossing locations} \label{app:future_nodes}

\begin{figure*}[!ht]
	\centering	
	\includegraphics[width=.99\textwidth]{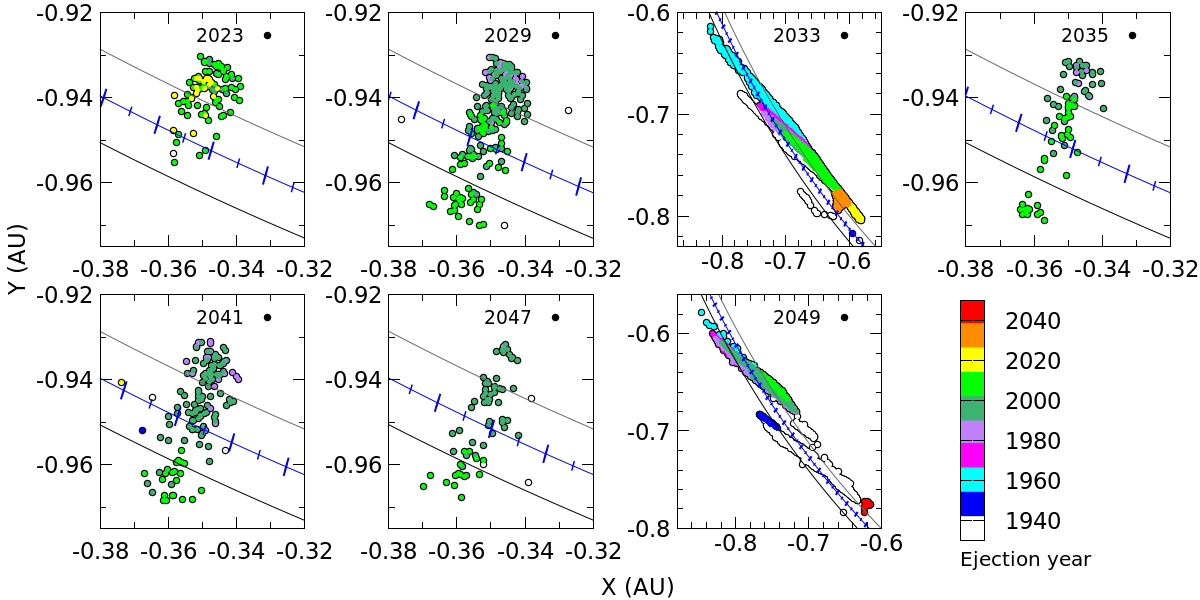}
	\includegraphics[width=.99\textwidth]{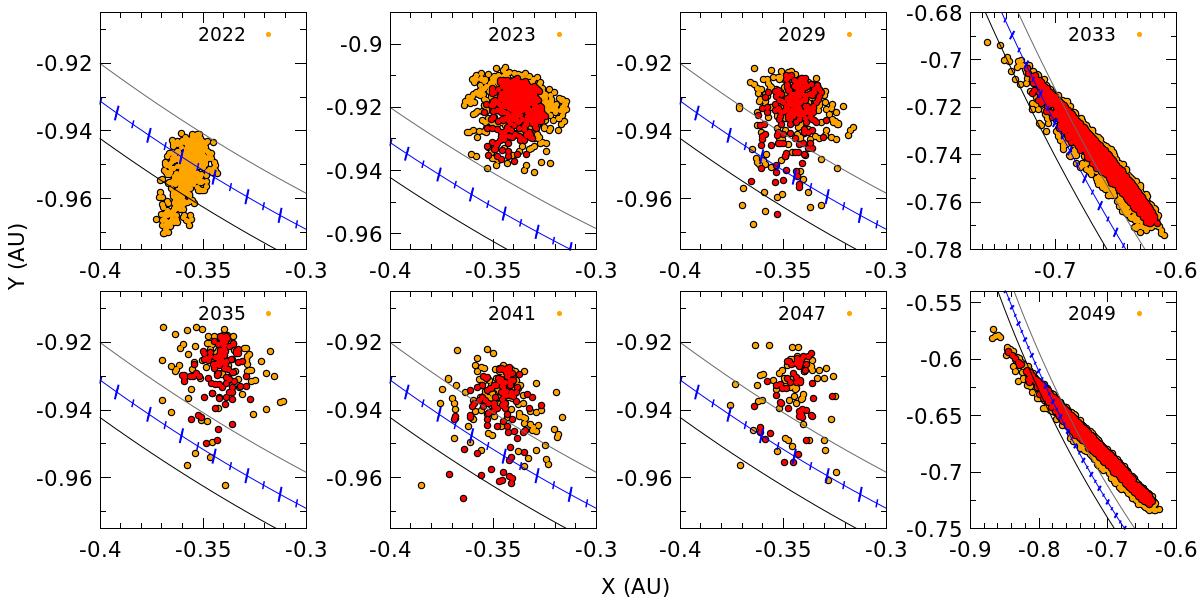}
	\caption{Nodal crossing location of the modelled TAH between 2023 and 2050. The top panel illustrates nodes of meteoroids ejected from comet 73P/73P-C since 1800, color-coded as a function of the ejection epoch. The bottom panels show the node evolution of meteoroids released during the 1995 break-up, with speeds of 1 (red) or 4 (orange) times the velocities predicted by the model of \cite{Crifo1997}. Only particles below $DX$=0.1 AU and $DT$=20 days from Earth's passage are represented. }
	\label{fig:future_nodes}
\end{figure*}

\clearpage
\newpage
\section{Predicted activity profiles} \label{app:predicted_profiles}

\begin{figure}[!ht]
	\centering
	\includegraphics[width=.49\textwidth]{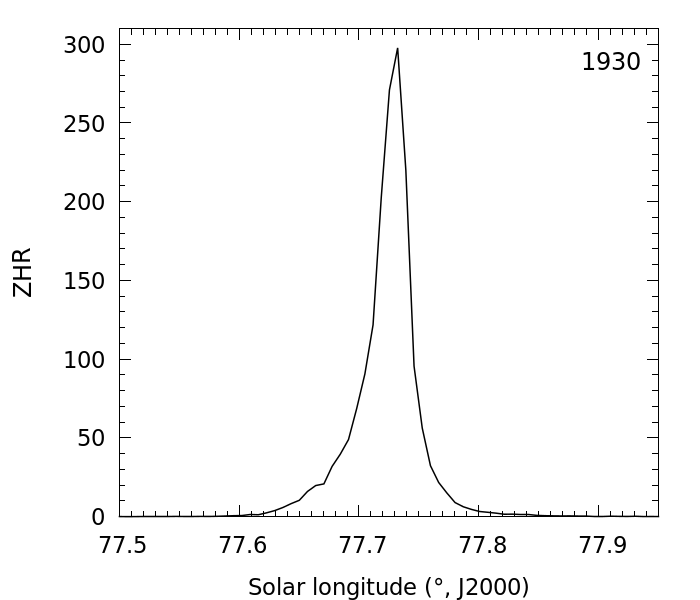}
	\includegraphics[width=.49\textwidth]{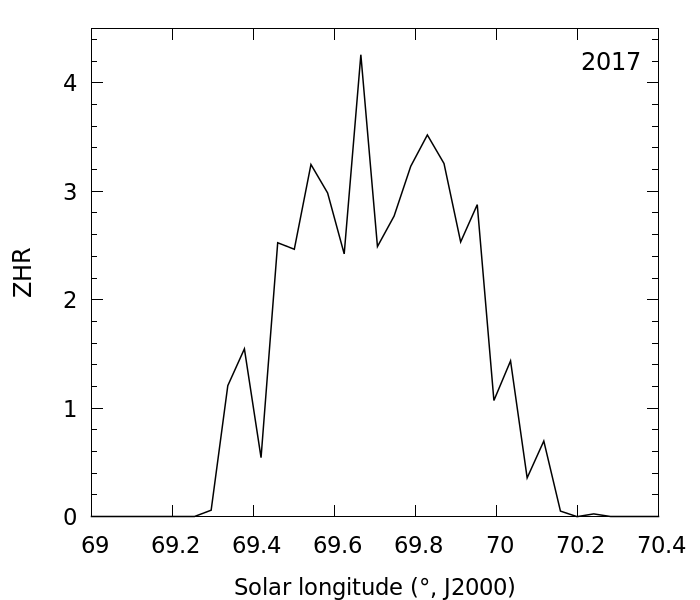}\\
	\caption{Simulated meteor activity in 1930 (left) and 2017 (right).  }
	\label{fig:1930_model}
\end{figure}

\begin{figure}[!ht]
	\centering
	\includegraphics[width=.49\textwidth]{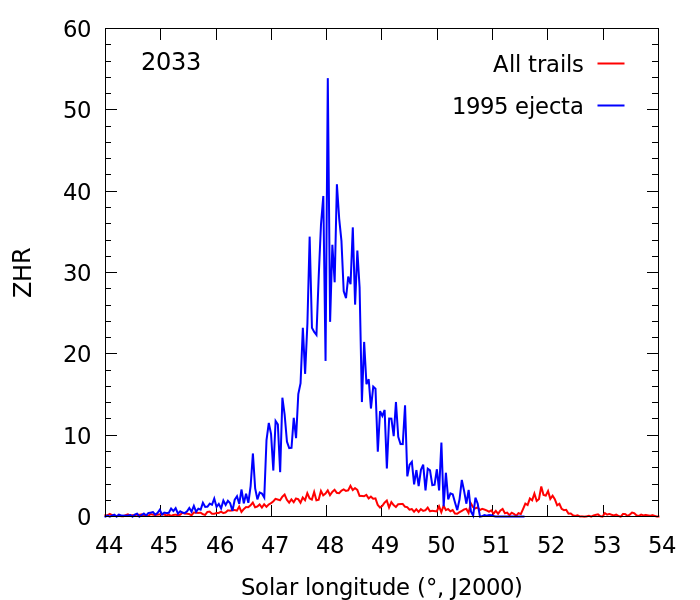}
	\includegraphics[width=.49\textwidth]{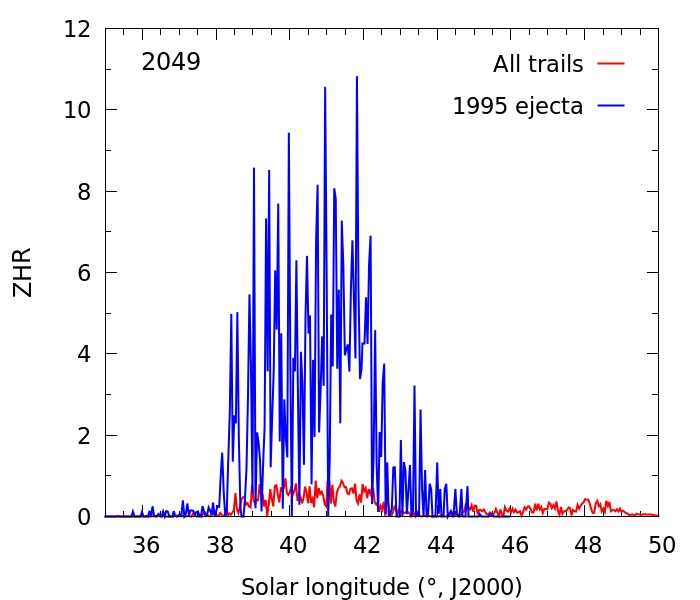}\\
	\caption{Simulated meteor activity in 2033 (left) and 2049 (right).  }
	\label{fig:future_profiles}
\end{figure}

\end{document}